\documentclass{aa}  

\usepackage{graphicx}
\usepackage{txfonts}
\usepackage{subcaption}

\usepackage{amsmath}
\usepackage{siunitx}
\usepackage{physics}

\usepackage[dvipsnames]{xcolor}
\definecolor{MyBlue}{HTML}{004DC9}
\definecolor{MyPurple}{HTML}{4E19B0}

\newcommand{\rep}[1]{\textcolor{black}{{#1}}}

\usepackage[linktocpage]{hyperref}
\hypersetup{
	colorlinks   = true,
	citecolor    = MyBlue, 
	urlcolor     = MyBlue,
	linkcolor    = MyPurple,
	breaklinks   = true,
	hypertexnames= true
}

\begin{document}
	
	\title{From photometric surveys to H\textsc{i} intensity mapping}
	
	\subtitle{Improving constraints on magnification biases while testing gravity}

	\author{Tiago Sinde\inst{1,2}\fnmsep\thanks{Corresponding author: up201906741@fc.up.pt}
		\and José Fonseca\inst{1,2,3}}
	
	\institute{Departamento de Física e Astronomia, Faculdade de Ciências, Universidade do Porto, Rua do Campo Alegre, 687, PT4169-007 Porto, Portugal
		\and Instituto de Astrofisica e Ci\^{e}ncias do Espa\c{c}o, Universidade do Porto CAUP, Rua das Estrelas, PT4150-762 Porto, Portugal
		\and Department of Physics \& Astronomy, University of the Western Cape, Cape Town 7535, South Africa}

	\abstract
	{The observed large-scale structure of the Universe is not a direct measure of the underlying distribution of matter. These observations are subtly distorted by gravitational lensing effects, which leave imprints on the statistical distribution of galaxies and offer a powerful test of general relativity.}
	{In this work, we investigate whether H\textsc{i} intensity mapping from current and forthcoming surveys can improve constraints on magnification lensing obtained from photometric galaxy surveys. In particular, we investigate whether one can jointly constrain the magnification bias parameters $s^\mathrm{G}(z)$ and the amplitude of the Weyl potential, which we parametrise as $\beta$.}
	{We employ a Fisher matrix formalism in order to estimate future constraints on the magnification biases and $\beta$. We forecast constraints for three photometric surveys (DES-like, LSST-like, Euclid-like) individually and together with two H\textsc{i} intensity mapping surveys (MeerKLASS, SKAO). We apply the multi-tracer technique by combining each galaxy survey with each H\textsc{i} survey, exploiting the combined constraining in the overlapping sky area.}
	{The multi-tracer approach dramatically improves constraints on $\beta$ by factors of \rep{18} to 50, depending on the surveys considered. For $s^\mathrm{G}(z)$, improvements can be marginal or by factors of 2 to \rep{4}. We also verify that $\beta$ and $s^\mathrm{G}(z)$ can be constrained simultaneously as the cross-correlations between tracers break the degeneracies among them.}
	{We conclude that the multi-tracer combination of photometric galaxy surveys and H\textsc{i} intensity mapping surveys enables high-precision measurements of both $s^\mathrm{G}(z)$ and $\beta$. This opens an additional pathway to constrain $\Phi+\Psi$ and test the validity of general relativity on cosmological scales.}
	

	\maketitle
	\nolinenumbers

	\section{Introduction} \label{sec:Intro}
		
		The three-dimensional large-scale structure (LSS) of the Universe encodes a wealth of information about its fundamental constituents, history, and underlying laws of gravity. Therefore, precision mapping of the distribution of matter in the Universe provides a powerful tool to constrain cosmological parameters and provide critical tests for models beyond the standard Lambda cold dark matter ($\Lambda$CDM) paradigm, including modifications to general relativity (GR).
		
		To trace dark matter, we use objects that we can observe and that are found in haloes of dark matter. We then look at how such objects fluctuate around an average. For galaxy surveys angular positions and redshifts are measured. The fact that we observe redshifts and do not measure distances directly makes our estimates of the clustering of matter offer contributions beyond the underlying dark matter fluctuations \citep{Challinor_2011, Bonvin_2011}. These additional effects that contribute to the observed number counts are the so-called GR effects. The first such effect, redshift space distortions (RSD), was already identified in the late 1980s \citep{1987MNRAS.227....1K}. 
				
		The second most significant GR effect is magnification lensing. Physically speaking, it accounts for the effects on the observed light that travelled across the Universe and passed through the gravitational field of matter. Weak gravitational lensing causes the observed images of galaxies to be distorted, in size and/or shape, compared to their true forms \citep{mathias_weak}. For clustering, the change in the angular size of a source is what determines whether a galaxy is to be magnified (or de-magnified) and be included in a galaxy sample (or not). 
		Several authors have quantified the importance of the lensing contribution in number counts \citep[see e.g.][]{Dio_2013,2021JCAP...04..055J}, how well it can be detected \citep{Alonso2015,Fonseca_2015,Montanari_2015,2021JCAP...04..055J}, and how neglecting such a term in the model would bias constraints in photometric surveys \citep{2022A&A...662A..93E} and spectroscopic surveys \citep{2024A&A...685A.167E,2021JCAP...12..004V}. For intensity mapping \citep{2019arXiv191104527F}, where we do not detect individual sources but instead the integrated emission of an emission line, the GR corrections take a simpler form \citep{Hall_2013} which does not include contributions from magnification lensing. Still, when cross-correlating galaxies with intensity maps in the foreground one neatly isolates the magnification lensing contribution \citep{2021JCAP...11..010V}. In the case of using gravitational wave events as tracers of dark matter, the GR corrections take yet another form \citep{Namikawa,2023JCAP...08..050F,2024JCAP...02..023B} and have been shown to provide additional constraining power on the magnification lensing term \citep{2025MNRAS.537.1912Z,2026MNRAS.547ag307Z}.
		
		\citet{Montanari_2015} go beyond and try to assess how well they can constrain the Weyl potential from which we construct the convergence (see \autoref{sec:Theory}). Identifying deviations of the Weyl potential would be a smoking gun for modified gravity \citep{2014PhRvD..89b4026B,2023PDU....3901151C} and is how models of modified gravity using shear lensing of photometric galaxy surveys are generally tested \citep{2020A&A...642A.191E}. 
		Moreover, \citet{2024NatCo..15.9295T} and \citet{2024PhRvL.133u1004G} constructed an estimator of the Weyl potential using both the angular power spectra of sources and galaxy-galaxy lensing, which they have shown to be yet another promising test of the Weyl potential. 
		
		Here we follow the approach of \citet{Montanari_2015}. It is important to note that the effect of convergence in the clustering of sources is modulated by the magnification bias \citep{Maartens_2021}. In general, one simply interprets it as the slope of the luminosity function at a given magnitude cut. In practice, galaxy catalogues are not built linearly, and we have to estimate such biases from the data \citep{Elvin_Poole_2023,Legnani_2026_DES_Y6}.

		Currently there is a wealth of photometric galaxy surveys --- the Dark Energy Survey (DES) \citep{2026arXiv260114559D}; the upcoming Legacy Survey of Space and Time (LSST) \citep{LSST_DarkEnergyScience_2021}; the Euclid mission \citep{2025A&A...697A...1E} --- that have emerged as central primary tools for mapping the LSS. Their great strength lies in efficiently covering immense volumes of the sky and in delivering high-statistics measurements of galaxy positions and shapes. However, this efficiency comes at a cost: the use of broadband photometry to estimate redshifts introduces significant photometric redshift (photo-$z$) uncertainties. These uncertainties smear information along the line of sight, degrading our three-dimensional view of the cosmic web.
		An ideal partner for photometric surveys is neutral hydrogen intensity mapping (H\textsc{i }IM) surveys, such as the one currently being done by the MeerKAT Large Area Synoptic Survey (MeerKLASS) \citep{MeerKAT_2016_M_Santos} and the one planned with the future Square Kilometre Array Observatory (SKAO) \citep{SKA_red_book_2020}. 
		H\textsc{i} IM surveys map the aggregate 21cm emission from cosmic neutral hydrogen without resolving individual galaxies. While they have lower angular resolution, they provide spectroscopic-precision redshift information along the line of sight and are also very efficient at covering vast cosmological volumes, thus complementing the weakness of photo-z surveys.
		However, it is important to note that the multi-tracer confines us to use the overlap sky area and redshift range. 
		
		This paper aims to forecast how precisely future optical and radio surveys can jointly constrain the magnification bias parameter $s^\mathrm{G}(z)$ and the amplitude of the Weyl potential (as a complementary probe of gravity), which we define as $\beta$ (see \autoref{sec:Weak_lens_Cosmology}). Although degenerate in principle, one can break such degeneracies with the multi-tracer technique \citep{Seljak_2009,Patrick_McDonald_2009, Hamaus_2011, Abramo_2013}. Hence, we use the Fisher matrix formalism to assess what improvement is given to photometric galaxy surveys by H\textsc{i} IM . 
		
		Following this introduction,  \autoref{sec:Theory} details the theoretical framework for weak magnification lensing and its impact on LSS measurements, along with the formal definition of the \rep{angular power spectrum}. 
		In \autoref{sec:Surveys} we describe the experimental specifications associated with the three photometric galaxy surveys (DES-like, LSST-like, Euclid-like) and the two H\textsc{i} IM surveys (MeerKLASS, SKAO) considered in this analysis.
		Next, in \autoref{sec:FIM} we develop the Fisher matrix formalism and present the relevant parameters to constrain the forecasts.
		We present and discuss our results in \autoref{sec:Results}.
		To finish, \autoref{sec:Conclusions} summarises our findings, compares them with other works, and suggests directions for future steps.
		Moreover, \rep{in order to validate our work we include a parallel analysis done with the angular correlation function in \autoref{ap:ACF}}. Additionally, appendices \ref{ap:Cov_M_C_ell} to \ref{ap:deriv_estimators} provide further details on the mathematical derivations made throughout this article.

	\section{Effect of magnification lensing in LSS estimators}\label{sec:Theory}	
	
		In this section, we review the theoretical framework of this paper. 
		We begin by introducing the weak lensing convergence and its connection to the gravitational potentials. 
		Then, we present the statistical estimator used to quantify the projected matter distribution, the angular power spectrum \rep{(denoted by $C_\ell$), as well as the relevant transfer functions.}
		
		Throughout this paper we consider the linearly perturbed flat Friedmann-Lemaître-Robertson-Walker (FLRW) metric in Newtonian gauge
		\begin{equation}\label{eq:metric}
			ds^2 = a^2(\eta) \left[ -(1+2\Psi) d\eta^2 + (1 - 2\Phi)\delta_{ij} dx^idx^j \right] \,.
		\end{equation}
		Here, $\Phi$ and $\Psi$ are the metric potentials, which are equal in $\Lambda$CDM and standard dark energy models, and $\eta$ denotes the conformal time.

		\subsection{Weak lensing convergence} \label{sec:Weak_lens_Cosmology}
	
			Gravitational lensing occurs when a body (or bodies) acts on a passing light ray as a lens, bending its trajectory due to gravity.
			This phenomenon can distort not only the observed position of a source but also its observed size and/or shape \citep{mathias_weak}.
			When the gravitational lensing system aligns in such a way that the lensing causes a focusing of light-rays towards the observer, the source's flux is amplified. This effect is known as magnification lensing. 
			This magnification, $\mu$, is related to lensing convergence $\kappa$ and complex shear $\gamma$ as \citep{Umetsu_2020}
			\begin{equation}
				\mu\equiv\frac{1}{(1-\kappa)^2-|\gamma|^2} \simeq 1 + 2\kappa \,. 
				\label{eq:magnification_lens}
			\end{equation}
			The convergence, $\kappa$, governs the isotropic stretching (size distortion) of the source images, while the shear, $\gamma$, describes their shape distortion.
			The approximation in \autoref{eq:magnification_lens} comes from the weak-lensing limit, where $|\gamma| \ll 1$ and $|\kappa| \ll 1$, in which we limit contributions to first order. This is the regime that we consider in our analysis.
			
			The lensing convergence observed in the direction $\hat n$ at redshift $z$ is given by an integral along the line of sight \citep{LEWIS_2006, Montanari_2015}
			\begin{equation}
				\kappa(\hat n, z) = \frac{1}{2}\int_0^{r(z)} \frac{r(z) - \tilde r}{r(z)\tilde r} \ \beta\ \nabla_\Omega^2\left( \rep{\Phi_{\Lambda CDM} + \Psi_{\Lambda CDM}} \right) \ d\tilde r \,,
				\label{eq:Mag_Conv_final}
			\end{equation}
			where $r$ is the comoving distance to the source. Here we follow \citep{Montanari_2015} and introduce the parameter $\beta$, which serves as a probe of modified gravity. In the standard $\Lambda$CDM model with negligible anisotropic stress from free-streaming neutrinos, one can take $\beta = 1$ \citep[see, e.g.][for a textbook discussion]{Baumann_2022}. The relation \rep{$\Phi_{\Lambda CDM} + \Psi_{\Lambda CDM}$} thus provides a direct null test of general relativity and \rep{since any modifications to gravity are absorbed into $\beta$, a value of $\beta \neq 1$ would signal a deviation from GR - or the presence of large anisotropic stress (just as a ratio of $\Phi_{\Lambda CDM}/\Psi_{\Lambda CDM} \neq 1$ would)}. 
			\rep{Consequently}, \autoref{eq:Mag_Conv_final} reveals a clear pathway for null testing gravity: \rep{ jointly constraining $\kappa$ and $\beta$ from observations, would grant direct access to $\Phi_{\Lambda CDM}+\Psi_{\Lambda CDM}$}. \rep{Note that, from now on we drop the $\Lambda CDM$ subscript from the potentials $\Phi$ and $\Psi$, even though they are still evaluated within standard models.}

		\subsection{\rep{Estimating} the clustering of matter} \label{sec:2D_estimators}

			The distribution of matter in the Universe extends throughout a three-dimensional space, making a three-dimensional statistical estimator of its distribution a natural option. In practice, we have to adapt the estimator to our observations. The well-known 3D matter power spectrum is an intuitive choice for studying its distribution, if we neglect redshift evolution. However, this method involves converting observed angular positions and redshifts into comoving distances, a process that relies on an assumed fiducial cosmological model. Although this is not a concern in spectroscopic surveys, due to the higher precision of redshift measurements and the Alcock-Paczynski test, it is not suited for photometric surveys where redshift measurements are less precise. 
			Fortunately, 2D statistics do not face this issue as they operate directly on the observed celestial sphere by only using the observed angular positions of galaxies. Although this method involves projecting all galaxies in a redshift bin, it is preferred for the scope of this work, as it provides better models for the clustering of photometric galaxies and includes the corrections from magnification lensing. 
			For these reasons, we intend to use the angular power spectrum, \rep{a 2D statistical estimator that ensures} our findings are independent of the underlying cosmology.
			
			To estimate the fluctuation of a galaxy sample we define the number density contrast 
			\begin{equation}
				\Delta_N(\hat n, z)=\frac{N(\hat n, z)-\langle N(\hat n, z)\rangle}{\langle N(\hat n, z)\rangle} \,,
			\end{equation}
			where $N(\hat n, z)$ is the number of galaxies in a certain pixel of the sky and $\langle N(\hat n, z)\rangle$ is its average. 
			Since this function lies on the surface of a sphere, we can decompose it into spherical harmonics
			\begin{equation}
				\Delta(\hat n, z) = \sum_{\ell=0}^{\ell} \sum_{m=-\ell}^{m=\ell} a_{\ell m}(z) Y_{\ell m}(\hat n) \,, \label{eq:Numb_counts_spher_harm} 
			\end{equation}
			where
			\begin{equation}
				a_{\ell m} =\int \Delta(\hat n, z) Y^*_{\ell m}(\hat n)  d\Omega \,.
			\end{equation}
			Here, $a_{\ell m}$ are the amplitude of the perturbations in the mode $Y_{\ell,m}(\hat n)$ on the surface of a sphere with radius $z$.
			
			For an unmasked surface, the statistical average of \(a_{\ell m}\) is \(\langle a_{\ell m} \rangle = 0\), implying that we cannot measure the amplitudes of the fluctuations of matter directly.
			Instead, we characterise their variance
			\begin{equation}
				\langle a_{\ell m}(\rep{z_i}) a_{\ell' m'}^*(\rep{z_j})\rangle \equiv \delta_{\ell\ell'}\delta_{mm'} C_\ell(\rep{z_i}, \rep{z_j}) \,.
				\label{eq:a_lm_C_ell}
			\end{equation}
			Here, \(C_\ell\) quantifies the variance of $a_{\ell m}$. A general expression for the angular power spectra of a bin $i$ with tracer $X$ and bin $j$ with tracer $X'$, is given by \citep{2026MNRAS.547ag307Z}
			\begin{equation}
				C^{\mathrm{W_{X}^{\rep{i}}} \mathrm{W_{X'}^{\rep{j}}}}_{\ell} = 4\pi \int \mathcal P(k) \Delta^{\mathrm{W_{X}^{\rep{i}}}}_{\ell}(k) \Delta^{\mathrm{W_{X'}^{\rep{j}}}}_{\ell}(k) \ \dd{\ln k} \,\,.
				\label{eq:C_ell_theo}
			\end{equation}
			The primordial matter power spectrum, $\mathcal P(k)$, is given by
			\begin{equation}
				\mathcal P(k) = A_s \left(\frac{k}{k_0}\right)^{n_s-1} \, ,
				\label{eq:Primordial_Pk}
			\end{equation}
			where $n_s$ is the spectral index, $A_s$ is the amplitude of primordial fluctuations, and the pivot scale $k_0 = 0.05 \text{Mpc}^{-1}$. The quantity $\Delta^{\mathrm{W_{X}^{\rep{i}}}}_{\ell}$ in \autoref{eq:C_ell_theo} is obtained in two steps. First, the number counts, from \autoref{eq:Numb_counts_spher_harm}, is Fourier transformed and expanded into $k$ and $\ell$ multipoles, $\Delta^X(\hat n, z) \rightarrow \Delta^X_\ell(z, k)$.
			Second, we account for the finite thickness of redshift bins by describing them through a window selection function $W(z, z_c)$ and a redshift distribution of sources $p^X(z)$ \citep{Dio_2013, Fonseca_2018} 
			\begin{equation}
				\Delta^{\mathrm{W_{X}^{\rep{i}}}}_{\ell}(k) \equiv \int p^X(z) W(z, z_i) \Delta^X_\ell(z, k) \ dz \,.
				\label{eq:Delta_W}
			\end{equation}
			To compute the angular power spectra we use the publicly available code \texttt{CAMB} \citep{Challinor_2011}.

		\subsection{Transfer functions}
	
			In \autoref{eq:Delta_W}, the expression of the transfer function for perturbations of a tracer $X$, $\Delta^{X}_{\ell}$, is composed of several physical contributions \citep{Fonseca_2016}
			\begin{align}
				\Delta^{X}_{\ell}(k) &= \Delta^{X,M}_{\ell} + \Delta^{X,RSD}_{\ell} + \Delta^{X,L}_{\ell}+\ldots \,,
				\label{eq:General_Frac_Pert_X_ell} 
			\end{align}
			where the superscripts denote matter density (M), redshift-space distortions (RSD), and lensing magnification (L). We have neglected Doppler, Sachs-Wolfe, and time-delay terms, as they have a negligible contribution to the signal. For simplicity, we have omitted the redshift dependence.
			
			The first term of \autoref{eq:General_Frac_Pert_X_ell} describes the intrinsic clustering of the tracer \citep{Challinor_2011, Bonvin_2011}
			\begin{align}
				\Delta^{X,M}_{\ell}(k) &= \Bigg[ b^{X}\delta^{S}_{k} -3\frac{\mathcal{H}v_{k}}{k}\Bigg] j_{\ell}(k r) \,, \label{eq:Delta_X_Matter} 
			\end{align}
			where $\delta^S$ is the dark matter density contrast in synchronous gauge, and $b^X(z, k)$ is the clustering bias of tracer $X$. For simplicity, we neglected the evolution bias. $\mathcal H$ is the conformal Hubble parameter, $r$ is the comoving line-of-sight distance, $v_k$ is the peculiar velocity, and $j_\ell$ is the spherical Bessel function of multipole $\ell$. 
			
			The second term of \autoref{eq:General_Frac_Pert_X_ell} is the contribution of the redshift-space distortion, which is independent of the chosen tracer (given the assumption that there is no velocity bias) \citep{1987MNRAS.227....1K}
			\begin{equation}
				\Delta^{X,RSD}_{\ell}(k) = \frac{k v_{k}}{\mathcal{H}} j''_{\ell}(k r)\,.
			\end{equation}
			
			The third term of \autoref{eq:General_Frac_Pert_X_ell} gives the contribution of lensing convergence to the tracer's fluctuations, integrated along the line of sight to each source \citep{Challinor_2011, Bonvin_2011,Montanari_2015}
			\begin{align}
				\Delta^{X,L}_{\ell}(k) &= \ell(\ell+1)\big(2-5s^{X}\big) \kappa_\ell   \label{eq:Delta_X_Lensing} \\
				&= \ell(\ell+1)\left(1-\frac{5}{2}s^{X}\right)
				\int_{0}^{r} \frac{r-\tilde r}{r\tilde r} \beta \left(\tilde\Phi_{k}+\tilde\Psi_{k}\right)\;j_{\ell}(k\tilde r) \  \dd\tilde r \,. \nonumber
			\end{align}
			The lensing effect amplitude is modulated by the $\beta$ parameter (see \autoref{eq:Mag_Conv_final}) and the magnification bias, $s^X$. 
			Here we need to make a careful distinction between number counts and intensity mapping. In linear order, there is no lensing contribution to the H\textsc{i} intensity fluctuations \citep{Hall_2013}. This follows from conservation of surface brightness in gravitational lensing. 
			For galaxy number counts, $s^\mathrm{G}$ is the logarithmic slope of the cumulative luminosity function $n^\mathrm{G} (z, m < m_\star )$ at the magnitude limit $m_\star$ of the survey. 
			Thus, we have \citep{Maartens_2021}
			\begin{align}
				s^\mathrm{G}(z) &= \frac{\partial\log_{10} n^\mathrm{G}(z, m < m_\star )}{\partial m_\star} \,.
			\end{align}

	\section{Surveys}\label{sec:Surveys}
	
		This section details the experimental specifications for the five surveys considered in our analysis. 
		We modelled three photometric galaxy surveys: a DES-like \citep{Legnani_2026_DES_Y6}, an LSST-like (Year 10) \citep{LSST_DarkEnergyScience_2021}, and a Euclid-like \citep{Euclid_2022}; and two H\textsc{i} intensity mapping surveys: MeerKLASS (UHF band) \citep{MeerKAT_2016_M_Santos} and SKAO (SKA1-MID Wide Band 1) \citep{SKA_red_book_2020}.
		For each survey, we specify the redshift distribution, window functions, biases, noise properties, and sky coverage required to compute the angular power spectra.

		\subsection{Photometric galaxy surveys} \label{sec:Galaxy_Surveys_Specifications}
	
\begin{figure}
				\centering
				\includegraphics[width=1.0\linewidth]{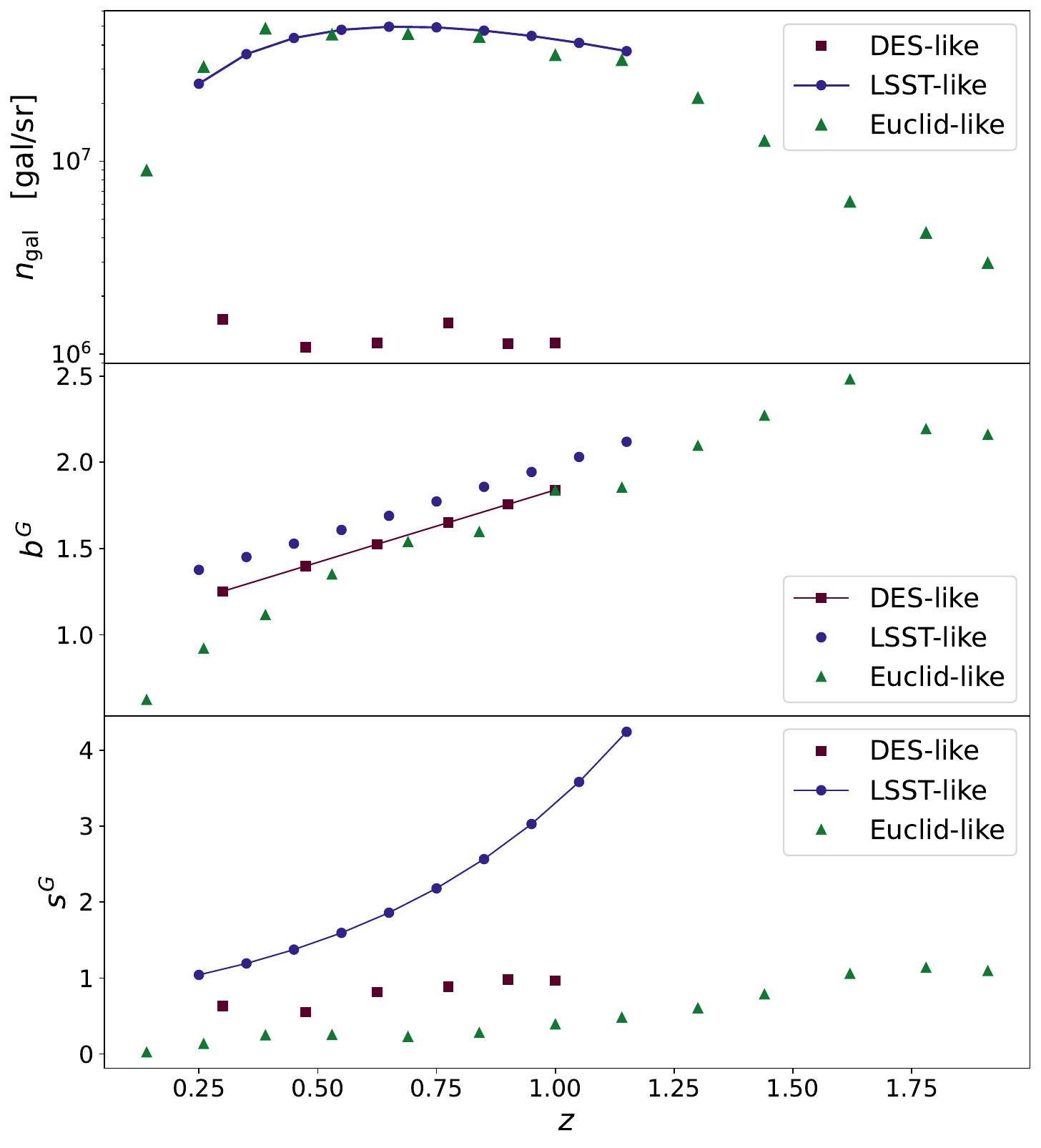}
				\caption{Values used for the number density of galaxies, $n_\mathrm{gal}$, the galaxy bias, $b^G(z)$,  and the magnification bias, $s^G(z)$ of the three galaxy surveys considered. Markers indicate measured or tabulated values; lines represent fitted or interpolated functions used in our analysis.}
				\label{fig:Galaxy_Surveys-bz_ngal_sz}
			\end{figure}

			For the three photometric galaxy surveys considered, we list in \autoref{tab:Surveys_Gal_dist} the redshift dependence of the distribution of sources $p(z)$ (see \autoref{eq:Delta_W}) that we assume. For the DES-like and Euclid-like surveys, we used the values of $n_\mathrm{gal}(z_i)$ listed in \citet{Legnani_2026_DES_Y6} and \citet{Euclid_2022}, respectively. For the LSST-like survey, we obtained $n_\mathrm{gal}(z)$ by imposing normalisation $n_\mathrm{LSST-like} \simeq 5.6 \times 10^8$~[gal/sr] \citep{LSST_DarkEnergyScience_2021} to the functional form listed in \autoref{tab:Surveys_Gal_dist}. In the top panel of \autoref{fig:Galaxy_Surveys-bz_ngal_sz} we plot the number densities we used in each redshift bin of the surveys considered.
			
			\begin{table}[ht!]
				\caption{\label{tab:Surveys_Gal_dist}Non-normalised probability distribution function of photometric galaxy surveys.}
				\centering
				\begin{tabular}{lc}
					\hline \hline 
					Survey & $p(z)$  \\
					\hline 
					& \\
					DES-like$^1$   & $ z^{1.04} \exp\left[-\left(\frac{z}{0.57}\right)^{1.34} \right]$  \\
					& \\
					LSST-like$^2$ & $ z^2\exp\left[-(\frac{z}{0.28})^{0.90}\right]$\\
					& \\
					Euclid-like$^3$ & $ z^2 \exp\left[ -(\frac{z}{0.79})^{1.5}\right]$  \\
					& \\
					\hline 
				\end{tabular}
				\tablebib{(1)~\citet{Fonseca_2016}; 
					(2)~\citet{LSST_DarkEnergyScience_2021}; 
					(3)~\citet{Montanari_2015}.
				}
			\end{table}
			
			Photometric surveys sacrifice redshift precision for survey speed, resulting in broad redshift distributions within each bin.
			The window function $W(z, z_i)$ in \autoref{eq:Delta_W} quantifies this effect, describing the probability that a galaxy at true redshift $z$ is assigned to the tomographic bin $i$ centred at $z_i$. 
			Following \citet{Ma_2006}, we model these windows using a combination of error functions
			\begin{equation}
				\begin{split}
					W(z, z_c; \Delta z_c, \sigma_{z_c}) = \, &\frac{1}{2} \Bigg\{ \erf\left[\frac{1}{\sqrt{2}\,\sigma_{z_c}} \left(z_c -z + \frac{\Delta z_c}{2}\right) \right] \\
					&\,\: -\erf\left[\frac{1}{\sqrt{2}\,\sigma_{z_c}} \left(z_c -z - \frac{\Delta z_c}{2}\right) \right] \Bigg\} \,.
				\end{split}
				\label{eq:Window_err_func}
			\end{equation}
			Here, $\Delta z_c$ is the bin width and $\sigma_{z_c} = \sigma_0(1+z_c)$ is the photometric redshift error.
			
\begin{figure}
				\centering
				\includegraphics[width=1.0\linewidth]{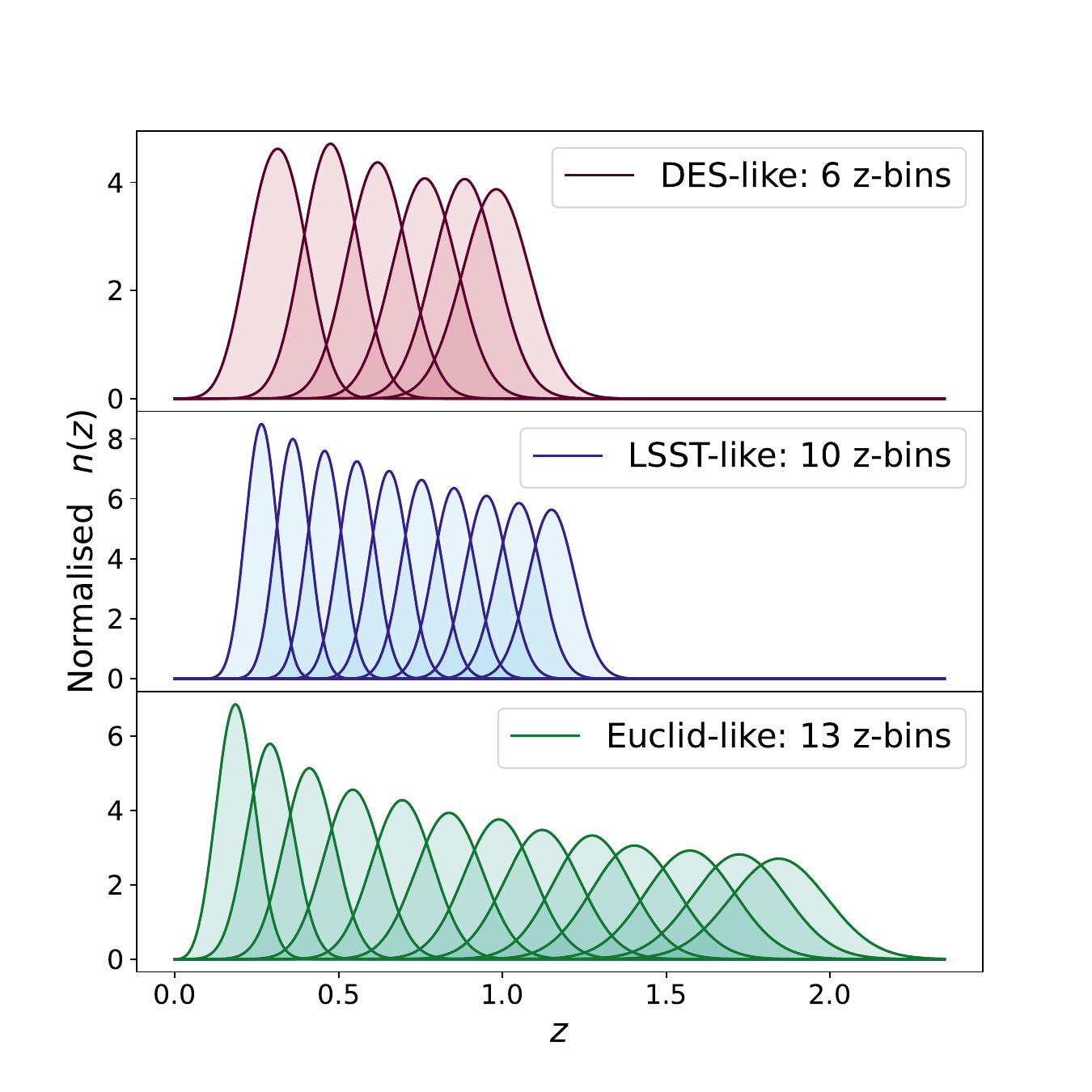}
				\caption{Redshift bins considered for the DES-like, LSST-like, and Euclid-like surveys. The values of the central redshifts of each bin are shown in \autoref{tab:z_c_values}.}
				\label{fig:Window_z-bins_Galaxies}
			\end{figure}

			So, in order to build these window functions, we start by defining the values of $z_c$, which are shown in \autoref{tab:z_c_values}.
			\begin{table}
				\caption{\label{tab:z_c_values}Central redshifts used for each bin of the considered surveys.}
				\centering
				\begin{tabular}{lccccc}
					\hline\hline
					$z_c$ & \small{DES-like} & \small{LSST-like} & \small{Euclid-like} & \small{MeerKLASS} & \small{SKAO} \\ 
					\hline
					$z_1$    & 0.30 & 0.25 &  0.14  & 0.50 & 0.40  \\
					$z_2$    & 0.48 & 0.35 &  0.26  & 0.70 & 0.60  \\
					$z_3$    & 0.63 & 0.45 &  0.39  & 0.90 & 0.80  \\
					$z_4$    & 0.78 & 0.55 &  0.53  & 1.10 & 1.00  \\
					$z_5$    & 0.90 & 0.65 &  0.69  & 1.30 & 1.20  \\
					$z_6$    & 1.00 & 0.75 &  0.84  &      & 1.40  \\
					$z_7$    &      & 0.85 &  1.00  &      & 1.65  \\
					$z_8$    &      & 0.95 &  1.14  &      & 1.95  \\
					$z_9$    &      & 1.05 &  1.30  &      &       \\
					$z_{10}$ &      & 1.15 &  1.44  &      &       \\
					$z_{11}$ &      &      &  1.62  &      &       \\
					$z_{12}$ &      &      &  1.78  &      &       \\
					$z_{13}$ &      &      &  1.91  &      &       \\
					\hline
				\end{tabular}
				\tablefoot{The bins for a MeerKLASS survey were chosen according to the z-range of the UHF band and the same was done for the SKAO survey with the same z-range as the MID Wide Band 1.}
				\tablebib{
					(\small{DES-like})~\citet{Legnani_2026_DES_Y6};
					(\small{LSST-like})~\citet{LSST_DarkEnergyScience_2021};
					(\small{Euclid-like})~\citet{Euclid_2022};
					(\small{MeerKLASS})~\citet{MeerKAT_2016_M_Santos};
					(\small{SKAO})~\citet{SKA_red_book_2020}
				}
			\end{table}			
			Then, for the DES-like and Euclid-like surveys, we considered that the width of the bins is the distance between two consecutive central redshifts, while their redshift scatter is $\sigma_0 = 0.05$.
			In the case of the LSST-like survey we used $\Delta z = 0.1$ and $\sigma_0 = 0.03$.
			The resultant window functions can be visualised in \autoref{fig:Window_z-bins_Galaxies}. 

			Following \citet{Alonso2015} and \citet{Fonseca_2016}, we adopted a simulation-based model \citep{Weinberg_2004} for the clustering bias of galaxies
			\begin{equation}
				b^\mathrm{G}(\rep{z_c}) = 1 + 0.84 \rep{z_c} \,,
			\end{equation}
			which is used in the DES-like survey. 
			On the other hand, for the LSST-like and the Euclid-like surveys we used the values from \citet{LSST_DarkEnergyScience_2021} and \citet{Euclid_2022}, respectively. These are shown in the middle panel of \autoref{fig:Galaxy_Surveys-bz_ngal_sz}, along with the predicted values of the DES-like survey. 
			
			The magnification bias $s^\mathrm{G}(z)$ for the LSST-like survey follows the polynomial fit of \citet{Fonseca_2016}
			\begin{equation*}
				s^\mathrm{G}(\rep{z_c})=0.132 + 0.259\rep{z_c} - 0.281\rep{z_c}^2 + 0.691\rep{z_c}^3 - 0.409\rep{z_c}^4 + 0.152\rep{z_c}^5 \,.
			\end{equation*}
			For the DES-like and Euclid-like surveys we adopted the values tabulated in \citet{Legnani_2026_DES_Y6} and \citet{Euclid_2022}, respectively. These, and the LSST-like values, can be seen in the bottom panel of \autoref{fig:Galaxy_Surveys-bz_ngal_sz}.

			The shot noise contribution to the galaxy angular power spectrum is given by 
			\begin{equation}
				\mathcal{N}^{\mathrm{G} \, \rep{ij}} = \frac{\delta^{\rep{ij}}}{n^{\rep{i}}} \,, 
				\label{eq:Shot_Noise_Gal}
			\end{equation}
			where $\delta^{\rep{ij}}$ is the Kronecker delta, ensuring that noise only contributes to autocorrelations, and $n^{\rep{i}}$ is the number density of galaxies (galaxies per steradian) inside each bin $z_i$.

		\subsection{H\textsc{i} intensity mapping surveys}
	
			H\textsc{i} IM differs fundamentally from galaxy surveys: instead of detecting individual sources, it measures the collective 21cm emission from unresolved neutral hydrogen.
			This affects the observed angular power spectrum through the telescope's beam and the nature of the H\textsc{i} signal, implying that we have to add a beam term that is non-existent in galaxy surveys. 
			For radio observatories with dishes, such as MeerKAT and SKAO, we can approximate the beam, to first order, by a Gaussian. 
			In harmonic space, this becomes \citep{Fonseca_2021}
			\begin{align*}
				B_\ell^{\rep{i}} &= \exp\left\{ -\frac{\ell(\ell+1)\,\theta_{\mathrm{FWHM}}^2(z_i)}{16\ln 2} \right\} \, ,
			\end{align*}
			where the full width at half maximum (FWHM) angular resolution is given by 
			\begin{equation}
				\theta_{\mathrm{FWHM}}(z_i) = 1.22\,\frac{\lambda_{\mathrm{ H\textsc{i}}}(1+z_i)}{D_{\rm dish}} \,.
			\end{equation}
			Here, $\lambda_{\mathrm{H\textsc{i}}} = 21\ \mathrm{cm}$ is the rest-frame wavelength of the 21cm line, and $D_{\mathrm{dish}}$ is the dish diameter: $13.5\ \mathrm{m}$ for MeerKAT \citep{MeerKAT_2016_M_Santos} and $15\ \mathrm{m}$ for SKAO \citep{SKA_red_book_2020}.
			Then, the beam modifies the observed angular power spectrum as \citep{Fonseca_2021}
			\begin{equation}
				\rep{C_{\ell}^{\mathrm{H\textsc{i}} \, ij} = B_\ell^{\rep{i}}B_\ell^{\rep{j}} \, C_{\ell}^{\mathrm{W_{H\textsc{i}}^i W^j_{H\textsc{i}}}}\,.}
			\end{equation}
			\rep{Similarly, the cross-correlation between a photometric galaxy survey and an H\textsc{i} IM is given by}
			\begin{equation}
				\rep{ 
					C_{\ell}^{\mathrm{G H\textsc{i}} \, ij} = B_\ell^{\rep{j}} \, C_{\ell}^{\mathrm{W_{G}^{\rep{i}}  W_{H\textsc{i}}^{\rep{j}}} } \,.
				}
			\end{equation}
			
\begin{figure}
				\centering
				\includegraphics[width=1.0\linewidth]{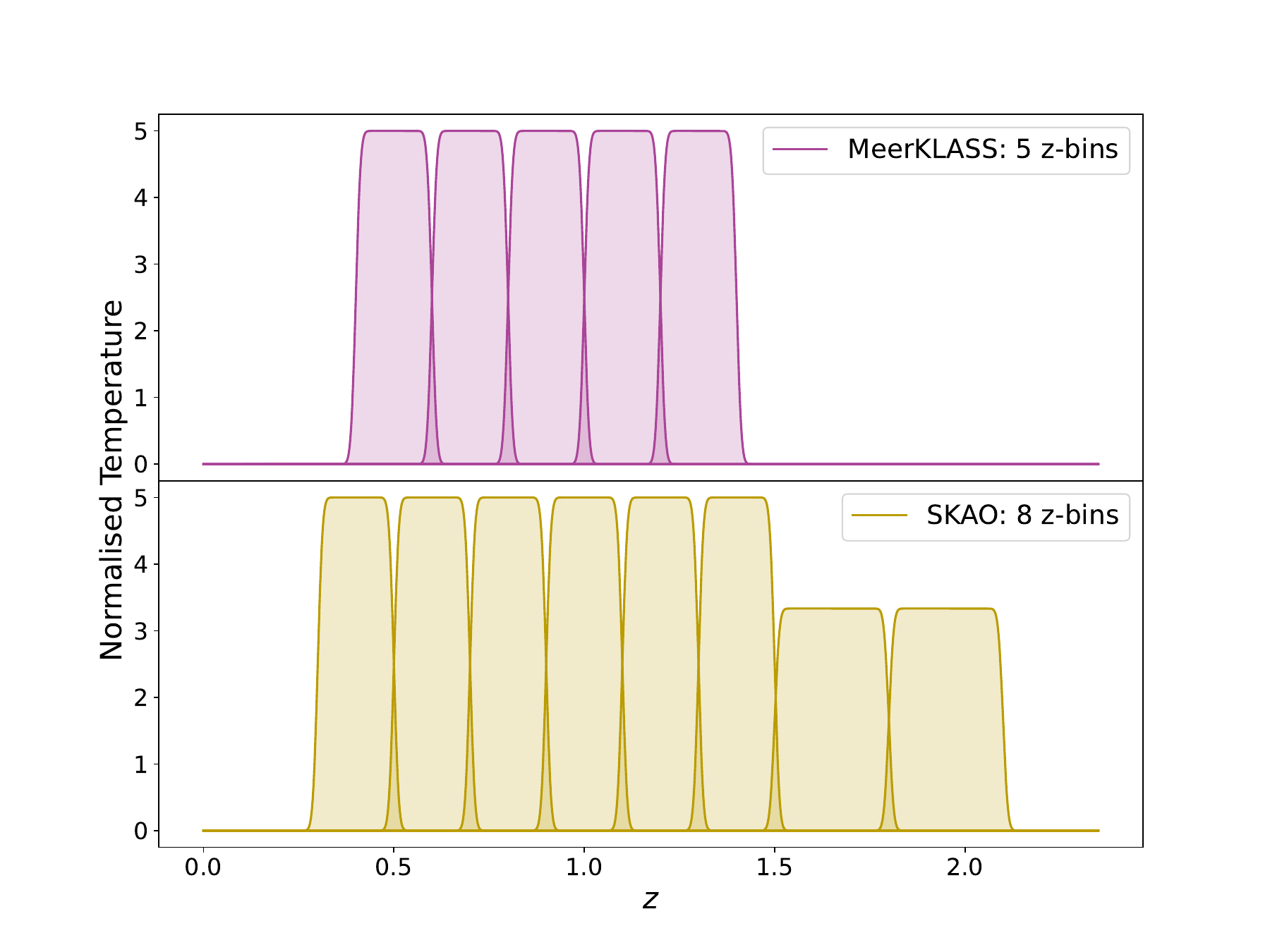}
				\caption{Redshift bins considered for the H\textsc{i} Surveys MeerKLASS and SKAO. The values of the central redshifts of each bin are shown in \autoref{tab:z_c_values}.}
				\label{fig:Window_z-bins_HI}
			\end{figure}

			In the calculation of $C_{\ell}^{\mathrm{W_{H\textsc{i}}^{\rep{i}} W_{X}^{\rep{j}}} }$ we have to define the H\textsc{i} IM redshift bins. In that case, the distribution of sources is only the H\textsc{i} temperature $T_\mathrm{H\textsc{i}}$, implying that $p^\mathrm{H\textsc{i}}(z) \propto T_\mathrm{H\textsc{i}}(z)$ in \autoref{eq:Delta_W}.
			Since the expression for the H\textsc{i} brightness temperature, $T_\mathrm{H\textsc{i}}$, at a certain redshift $z$  is poorly constrained by current observations, we used the fit from \citet{MeerKAT_2016_M_Santos}
			\begin{equation}
				\bar T_\mathrm{H\textsc{i}} (z_c) = 0.056 + 0.232z_c - 0.024z_c^2 \ [\mathrm{mK}] \, .
			\end{equation}
			Moreover, we \rep{utilised} the same window function formalism as in the galaxy surveys (\autoref{eq:Window_err_func}), with bin widths of $\Delta z = 0.2$ for $z_c < 1.5$ and $\Delta z = 0.3$ for $z_c > 1.5$, and a redshift scatter set to \rep{$\sigma_{z_c} = 0.01$}.
			\autoref{tab:z_c_values} lists the central redshifts of each bin, and \autoref{fig:Window_z-bins_HI} shows the resulting window functions.
			
\begin{figure}
				\centering
				\includegraphics[width=0.95\linewidth]{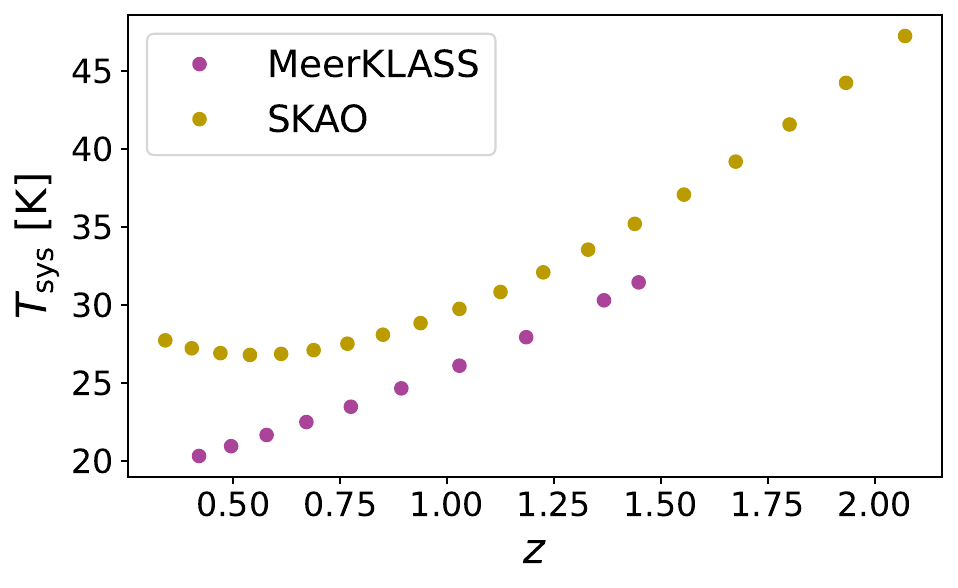}
				\caption{Values of the system temperatures, $T_\mathrm{sys}$, used in the analysis that includes H\textsc{i} IM surveys.}
				\label{fig:HI_Surveys-Tsys}
			\end{figure}

			For the H\textsc{i} clustering bias, $b^\mathrm{H\textsc{i}}$, we followed \citet{SKA_red_book_2020} and adopted the parametrisation 
			\begin{equation}
				b^\mathrm{H\textsc{i}}(\rep{z_c}) = 0.667 + 0.178\rep{z_c} + 0.050\rep{z_c}^2 \,.
				\label{eq:b_HI}
			\end{equation}
			In H\textsc{i} IM, the shot-noise is negligible, so the main noise contribution comes from the instrumental noise, i.e. the uncertainty in measuring the temperature in each pixel of the map. 
			The power spectrum of the instrumental noise reads \citep{Fonseca_2018}
			\begin{equation}
				\mathcal N^{\mathrm{H\textsc{i}} \, \rep{ij}} = \delta^{\rep{ij}} \frac{4\pi f_{\text{sky}}\left(T_{sys}^{\rep{i}}\right)^2}{2 N_d t_{tot} \Delta\nu^{\mathrm{H\textsc{i}} \, \rep{i}} }\,,
				\label{eq:Shot_Noise_HI}
			\end{equation}
			where $T_{sys}^{\rep{i}}$ is the system temperature for bin $i$, $\Delta\nu^{\mathrm{H\textsc{i}} \, \rep{i}}$ is the frequency width of the H\textsc{i} bin (in Hz), $N_d$ the number of collecting dishes, and	$t_{tot}$ the total observation time (in seconds).
			For MeerKLASS's UHF band, we considered $N_{\mathrm{d}} = 64$ and an updated integration time of $t_{\mathrm{tot}} = 2500$~hours for an updated survey area of $10000\deg^2$ \citep{MeerKLASS_UHF_2025}. 
			For SKAO Wide Band 1, we used $N_{\mathrm{d}} = 197$ and $t_{\mathrm{tot}} = 10000$~hours covering an area of $20000\deg^2$ \citep{SKA_red_book_2020}. 
			The system temperatures used for each redshift bin are interpolated from \autoref{fig:HI_Surveys-Tsys} by using the nearest available temperature to the central value of a certain z-bin.
			Observational surveys cover only a fraction of the sky, which we denote by $f_{\mathrm{sky}}$. 
			Since multi-tracer analyses combine two surveys, the effective sky fraction is the overlap area. What this means in practice is that when we combine two redshift bins from surveys with different values of $f_{\text{sky}}$ we always consider the smaller value of the two. 
			In \autoref{tab:sky_fractions} we summarise the sky fractions for each survey and the overlap fractions used in our forecasts.
			\begin{table}[ht!]
				\caption{\label{tab:sky_fractions}Sky coverage of each survey.}
				\centering
				\begin{tabular}{l c c}
					\hline
					\hline
					Survey & Area [deg$^2$] & $f_{\mathrm{sky}}$ \\
					\hline
					DES-like$^1$ & 5000 & 0.12 \\
					LSST-like$^2$ & 14300 & 0.35 \\
					Euclid-like$^3$ & 14000 & 0.34 \\
					MeerKLASS$^4$ & 10000 & 0.24 \\
					SKAO$^5$ & 20000 & 0.49 \\
					\hline
				\end{tabular}
				\tablefoot{
					The sky fraction $f_{\mathrm{sky}}$ is computed relative to full sky ($4\pi$[sr]). 
					For multi-tracer combinations, we use the overlapping sky area, corresponding to the smaller $f_{\mathrm{sky}}$ of the two surveys.}
				\tablebib{
					(1)~\citet{Legnani_2026_DES_Y6};
					(2)~\citet{LSST_DarkEnergyScience_2021};
					(3)~\citet{2025A&A...697A...1E};
					(4)~\citet{MeerKLASS_UHF_2025};
					(5)~\citet{SKA_red_book_2020}.
				}
			\end{table}

	\section{Fisher analysis} \label{sec:FIM}
	
		The Fisher information matrix provides a powerful tool for forecasting the precision with which a certain statistical estimator can measure a set of parameters \citep{Huterer_2023}. 
		In the context of this work, it allows us to quantify the minimum achievable observational errors on parameters $\lambda$ when \rep{considering} the angular power spectrum, $C_\ell$, as our statistical estimator. 
		The usefulness of Fisher matrix calculations is that, given some data $D$, it provides an essentially instantaneous result (compared to the long computational times of a Markov chain Monte Carlo (MCMC) simulation), without stochastic noise \citep{Huterer_2023}. 
		
		The expression for the Fisher matrix of either $C_\ell$ can be approximated by
		\begin{equation}
			F_{\mu\nu} \simeq \bar{\mathbf{D}}_{,\mu}^T \mathbf C^{-1} \bar{\mathbf{D}}_{,\nu} 	 \,,				
			\label{eq:FisherM_general}
		\end{equation}
		where $\bar{\mathbf{D}}$ is the mean of the data vector $\mathbf{D}$. In our analysis, this represents a vector of the angular power spectrum, \rep{$\mathbf{D} = \vec C_{\ell}^{ij}=\{\vec C_{\ell}^{\mathrm{G}}, \vec C_{\ell}^{\mathrm{H\textsc{i}}}, \vec C_{\ell}^{\mathrm{G H\textsc{i}}}\}$}.
		The vector $\vec C_{\ell}^{\rep{ij}}$ contains a value for each $\ell$ element that we are considering in a certain redshift bin combination, \rep{$i$ and $j$}. 
		
		In order to obtain a Fisher matrix we need the mean of the data vector and the covariance matrix of the data.
		For $C_\ell$, the mean of the data vectors are themselves, so $\bar{\mathbf{D}} = \langle \vec{C}_{\ell}^{\rep{ij}}\rangle = \vec{C}_{\ell}^{\rep{ij}}$.
		Then, its covariance matrix is denoted by
		\begin{align}
			\mathbf C = \text{Cov}\left( \vec{C}_{\ell}^{\rep{ij}},  \vec{C}_{\ell'}^{\rep{pq}}\right) \equiv \rep{\Gamma_{\ell\ell'}^{ijpq}} \,,
		\end{align}
		for which its expression was deduced in \autoref{ap:Cov_M_C_ell} and is given by \autoref{eq:CovM_Cell_noisy_lstep_ap}.
		
		Moreover, before computing the Fisher matrices, we still need to determine the derivative of $C_\ell$ in relation to a certain parameter $\lambda$. 
		In \autoref{ap:deriv_estimators} we outline how we used the five-point stencil approximation and two analytic derivatives to obtain the values of $\bar{\mathbf{D}}_{,\mu} =\partial \vec{C}_{\ell}^{\rep{ij}} / \partial \lambda_\mu$.
		
		Substituting by more explicit notation, we can rewrite the Fisher matrix expression of the angular power spectrum as
		\begin{align}
			F_{\mu\nu} 
			\simeq \left( \frac{\partial \vec{C}_{\ell}^{\rep{ij}}}{\partial \lambda_\mu}\right)^T \left(\rep{\Gamma_{\ell\ell'} ^{ijpq}}\right)^{-1} \frac{\partial \vec{C}_{\ell'}^{\rep{pq}}}{\partial \lambda_\nu} \,.
			\label{eq:Fish_Cl}	
		\end{align}
		
		Afterwards, we can use the inverse of the Fisher matrix to provide the lower bound on the variance of the parameters considered. 
		Then, the marginalised error of a certain parameter $\lambda$ is given by the Cramér-Rao bound \citep{heavens2010statisticaltechniquescosmology,Huterer_2023},
		\begin{equation}
			\sigma({\lambda_\mu}) \ge \sqrt{ (F^{-1})_{\mu\mu} }\,.
			\label{eq:min_err_FishM}
		\end{equation}

		In this work, we define the cosmology of a fiducial Universe according to the results from \citet{Planck2018}. 
		The fiducial values used for each cosmological parameter are defined in \autoref{tab:Cosm_param}. 			
		\begin{table}
			\caption{\label{tab:Cosm_param}Analysed cosmological parameter values.}
			\centering
			\begin{tabular}{cc}
				\hline\hline
				Cosmological Parameter & Fiducial  values \\ 
				\hline
				$H_0 $  & $67.5$ \\
				$w$ & $-1$ \\
				$\Omega_{cdm, 0}$ & $0.27$ \\
				$\Omega_{b,0}$ & $0.045$ \\
				$A_s$ & $2\times 10^{-9}$\\
				$n_s$ & $0.965$ \\		
				\hline
			\end{tabular}
			\tablebib{(1)~\citet{Planck2018}.}
		\end{table}
		Some of these parameters have already been mentioned ($A_S$ and $n_S$ in \autoref{eq:Primordial_Pk}), but we have not yet referred to most of them; $w$ is the dark energy equation of state parameter, while $\Omega_{cdm,0}$ and $\Omega_{b,0}$ are the cold dark matter and baryon density parameters, which are related to the density parameter for matter as $\Omega_{m,0} = \Omega_{cdm,0} + \Omega_{b,0}$.				
		Finally, $H_0$ is the present value of the Hubble parameter.
		In order to constrain the metric potentials we have to constrain the values of parameter $\beta$, and the magnification convergence $\kappa$; see \autoref{eq:Mag_Conv_final}.
		However, since we cannot constrain the value of $\kappa$ directly, we must first constrain $s^\mathrm{G}(z)$ and then take advantage of \autoref{eq:Delta_X_Lensing} to constrain $\kappa$.
		The fiducial value of $\beta$ is $1$, while the fiducials of $s^\mathrm{G}(z)$ are given in \autoref{fig:Galaxy_Surveys-bz_ngal_sz}.
		In summary, the vector containing all the parameters that we analyse is
		\begin{equation}
			\vec \lambda = \{ H_0, w, \Omega_b, \Omega_{cdm}, \ln A_s, n_s, \beta, s^\mathrm{G}(z_i) \} \,.
			\label{eq:lambda_parameters}
		\end{equation}
		\rep{Additionally, we note that no additional priors were included in our analysis. Although such priors could potentially improve the constraints, we found that the parameters $\beta$ and $s^\mathrm{G}$ are not so sensitive to them. Furthermore, investigating the relationship between these parameters and cosmological parameters was not the primary focus of this work.}
		
		\rep{Equipped with all the necessary components, we proceeded to the numerical implementation of these theoretical expressions in Python. In this way we computed the Fisher matrix forecasts and thereby evaluated the performance of the $C_\ell$ estimator.}
		
		To streamline the computation, we redefined our data vector as $\mathbf{D} = \vec C_{\ell}^{\rep{I}}$ where the index $I=\{ z_i z_j\}$ denotes a specific combination of redshift bins without repetition. This formulation yields longer data vectors that contain every value for each bin combination $I$ in all multipoles $\ell$.
		In our analysis, we included all possible bin combinations, incorporating both the autocorrelations and the cross-correlations between different redshift bins. 
		
		For a specific bin combination, \(I\), the Fisher matrix is calculated as follows:
		\begin{enumerate}
			\item Covariance matrices: Compute the covariance matrix of $\vec C_{\ell}$ with dimensions $I\ell \times I\ell$ following the methods in \autoref{ap:Cov_M_C_ell}.
			
			\item Parameter derivatives: Calculate the derivatives of the data vector $\partial \vec C_{\ell} / \partial \lambda_\mu$ as outlined in  \autoref{ap:deriv_estimators}. This vector has dimensions of $I\ell$.				
			
			\item Matrix inversion: Invert the covariance matrix, $\Gamma \rightarrow \Gamma^{-1}$, and transpose the derivatives as $\bar{\mathbf{D}}_{,\mu} \rightarrow \bar{\mathbf{D}}_{,\mu}^T$.
			
			\item Fisher matrix construction: Assemble the Fisher matrix using \autoref{eq:Fish_Cl}.
		\end{enumerate}			
		
		To interpret the forecasted constraints, we visualised the results using error ellipses and histograms of relative uncertainties rather than presenting raw numerical values. The relative uncertainty for a parameter $\lambda_\mu$ is defined as $\sigma(\lambda_\mu)/\lambda_\mu \times 100\%$, where $\sigma(\lambda_\mu)$ is the marginalised error obtained from the Fisher matrix.

	\section{Results} \label{sec:Results}
	
		\subsection{Degeneracies between $s^\mathrm{G}$ and $\beta$}
	
\begin{figure}
				\centering
				\includegraphics[width=1.0\linewidth]{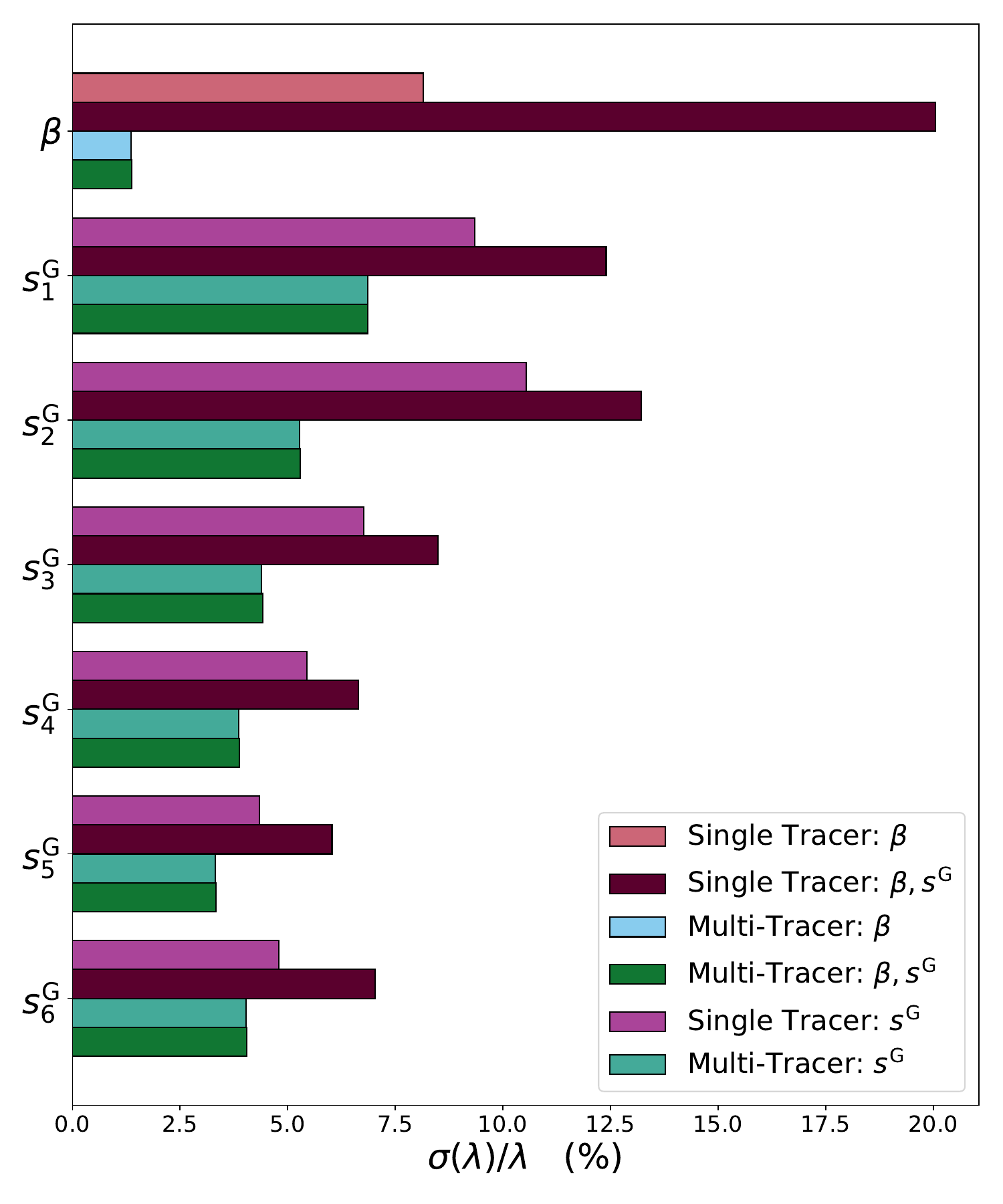}
				\caption{Comparison of parameter constraints obtained with different analysis choices (separate vs joint estimation). For the single tracer we used the DES-like survey and the multi-tracer denotes the combination of DES-like with the MeerKLASS survey.}
				\label{fig:Histogram_Compare_results_sz_beta_DES_MK}
			\end{figure}

			\rep{Before studying the impact of the multi-tracer technique, it is important to} analyse the relation between $s^\mathrm{G}$ and $\beta$.
			To achieve this, we compared two approaches: constraining $\beta$ and $s^\mathrm{G}(z)$ separately versus simultaneously. 
			The results are shown in the histogram in  \autoref{fig:Histogram_Compare_results_sz_beta_DES_MK}.
			In it we notice that there is an increase in relative uncertainty from constraining only one of the parameters, $s^\mathrm{G}(z_i)$ or $\beta$, to constraining them together.
			This difference comes from the fact that these parameters are locally degenerate, i.e. are degenerate for the same redshift bin and its autocorrelation. 
			However, this degeneracy is loosened when we include tomographic information through cross-correlations between different redshift bins.
			As shown in the figure, the addition of multiple H\textsc{i} intensity mapping redshift bins reduces the difference between separate and joint estimation to almost negligible levels, demonstrating the power of multi-tracer tomography to overcome parameter degeneracies. It is worth emphasising that the multi-tracer approach has a significantly greater impact on $\beta$, as it reduces the relative error from $20\%$ to approximately $1\%$.
			
		\subsection{Fisher forecasts: Single versus multi-tracer method} \label{sec:FIM_Multi-tracer}
	
\begin{figure}
				\centering
				\includegraphics[width=1\linewidth]{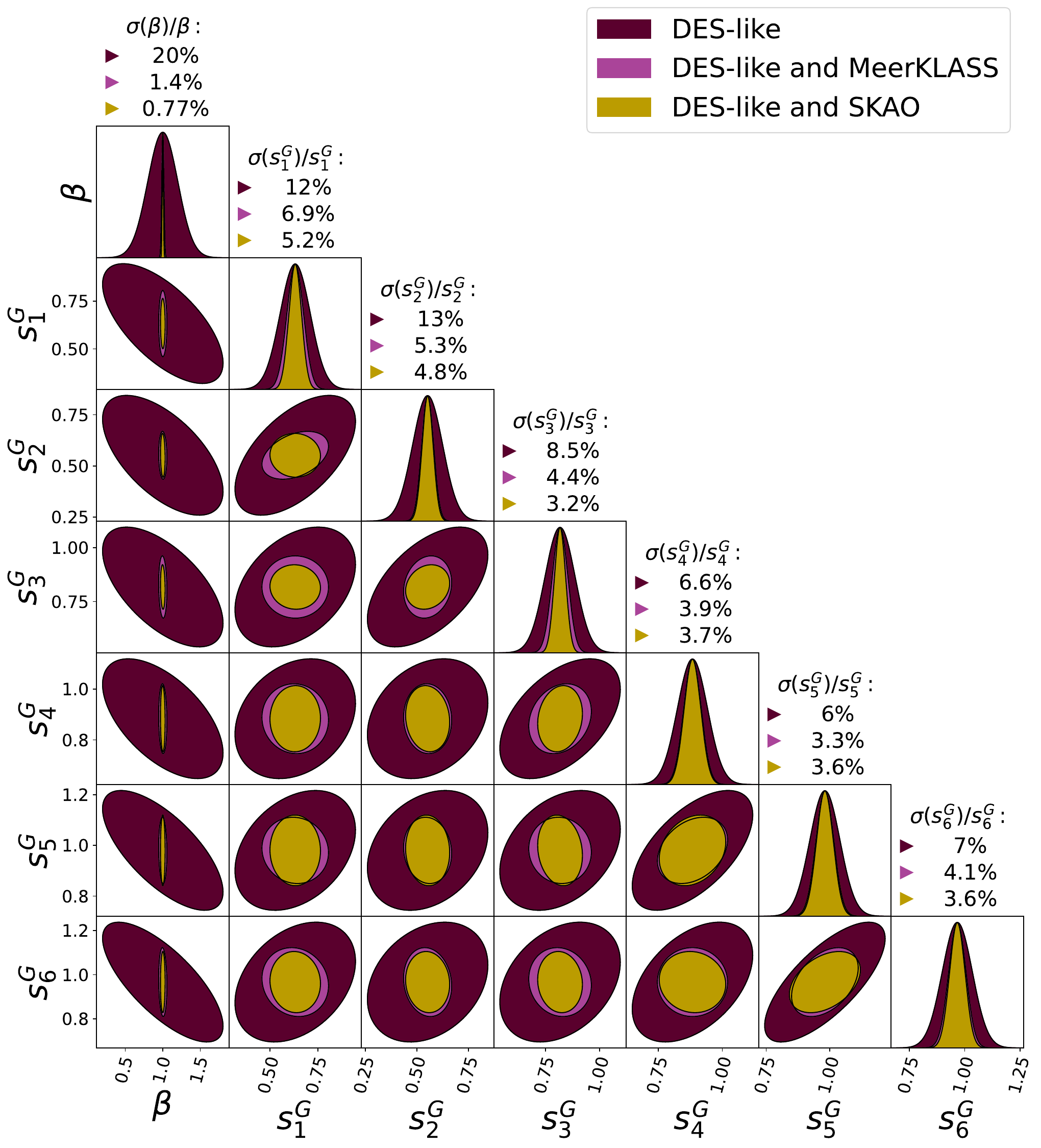}
				\caption{Corner plot showing constraints on $\beta$ and $s^\mathrm{G}(z_i)$ from the combination of DES-like and H\textsc{i} surveys (MeerKLASS and SKAO), compared to DES-like alone.}
				\label{fig:CornerP_DESxHI}
			\end{figure}

\begin{figure}
				\centering
				\includegraphics[width=1\linewidth]{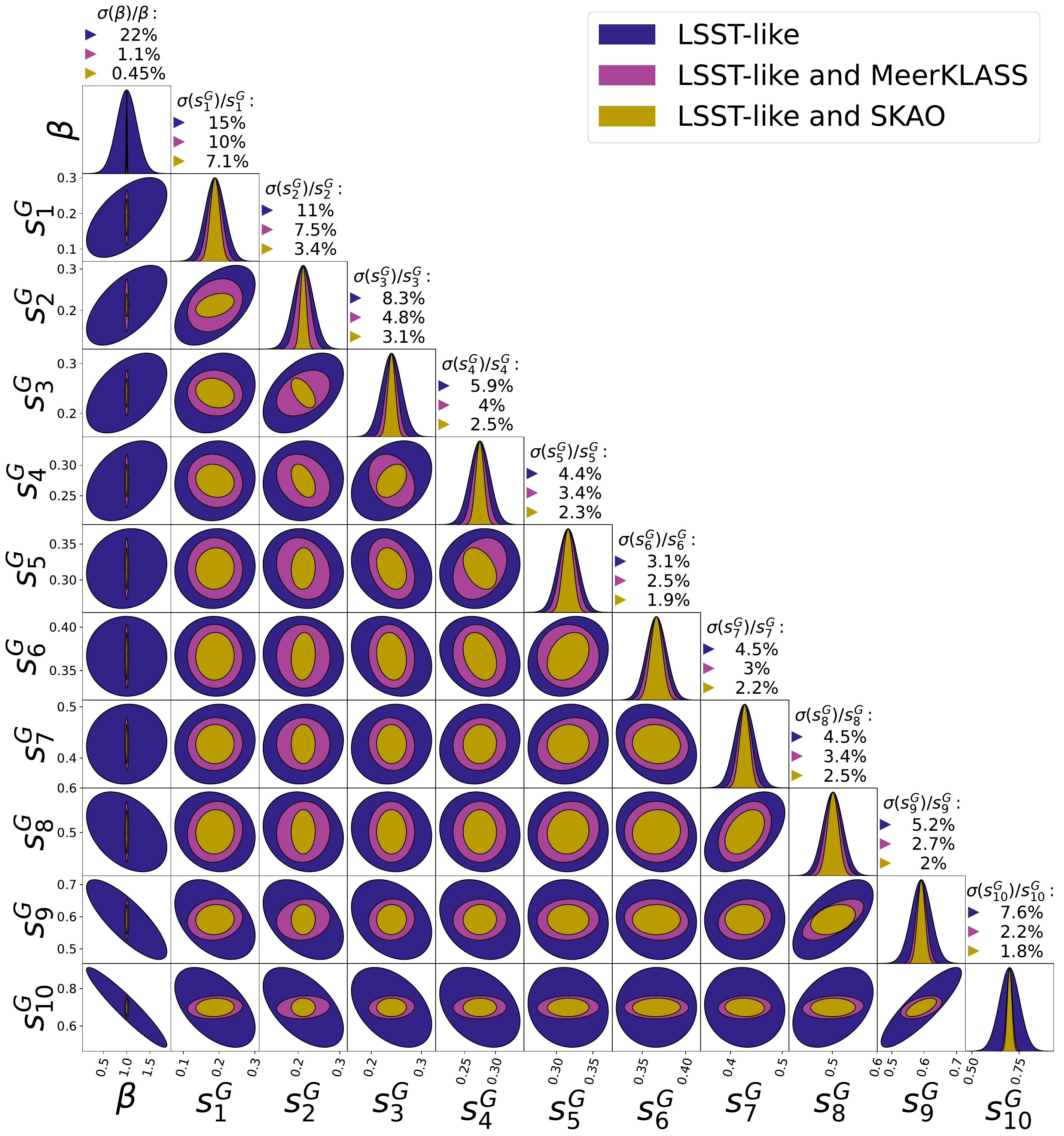}
				\caption{Same as \autoref{fig:CornerP_DESxHI} but for the combination of LSST-like with H\textsc{i} surveys.}
				\label{fig:CornerP_LSSTxHI}
			\end{figure}

\begin{figure}
				\centering
				\includegraphics[width=1\linewidth]{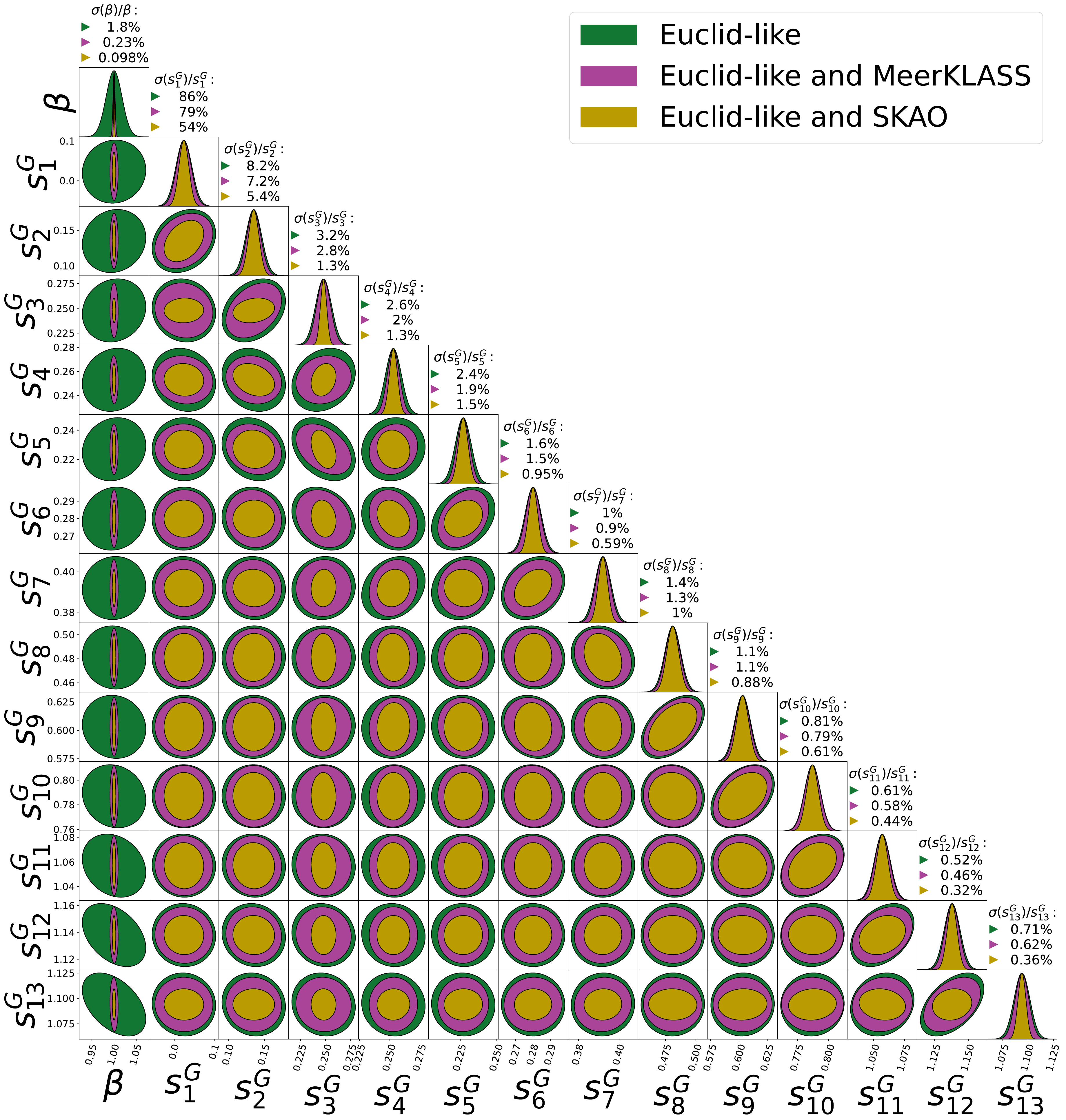}
				\caption{Same as \autoref{fig:CornerP_DESxHI} but for the combination of Euclid-like with H\textsc{i} surveys.}
				\label{fig:CornerP_EuclidxHI}
			\end{figure}

			In this section we present Fisher matrix forecasts using the multi-tracer technique to constrain the $\lambda_\mu$ parameters defined in \autoref{eq:lambda_parameters}. 
			For each galaxy survey, we compared the constraints on $\beta$ and $s^\mathrm{G}(z)$ for three scenarios: using the galaxy survey data alone, combining it with MeerKLASS, and combining with the SKAO data.
			The results are shown in \autoref{fig:CornerP_DESxHI} (DES-like), \autoref{fig:CornerP_LSSTxHI} (LSST-like), and \autoref{fig:CornerP_EuclidxHI} (Euclid-like).
			
			Our results demonstrate a substantial reduction in parameter uncertainties when combining galaxy surveys with H\textsc{i} intensity mapping surveys, further validating the effectiveness of the multi-tracer method.
			The improvement scales with the amount of spatial (galaxy distribution) information available: constraints become progressively tighter as we move from the six redshift bins of DES-like to the ten bins of LSST-like and finally to the 13 bins of Euclid-like, for both single and multi-tracer approaches.
			
			The most dramatic improvement is observed for parameter $\beta$, which probes modifications in general relativity. For the DES-like and LSST-like surveys, the uncertainty in $\beta$ decreases from approximately $20\%$ in the single-tracer case to less than $1\%$ when combined with H\textsc{i} surveys (see \autoref{fig:CornerP_DESxHI} and \autoref{fig:CornerP_LSSTxHI}). 
			The tightest constraints are achieved with the Euclid-like and SKAO combination, reaching a remarkable relative uncertainty of $\sim\rep{0.1}\%$ (see \autoref{fig:CornerP_EuclidxHI}).  
			
			For the magnification bias parameters $s^\mathrm{G}(z_i)$, the multi-tracer method also delivers improvements, though less dramatic than for $\beta$. 
			The best performance is again achieved with Euclid-like combined with SKAO, yielding uncertainties as low as $\sim\rep{0.35}\%$. 
			While our best results showed an improvement factor for $s^\mathrm{G}(z)$ between 2 and \rep{4} times better than for the single tracer case, the constraints on $\beta$ improved by factors of \rep{18} to 50. 
			Consequently, both sets of constraints are sufficiently precise to enable meaningful measurements of these parameters and, through them, access to the gravitational potentials $\Phi$ and $\Psi$.

			One noteworthy exception occurs for the fifth redshift bin of DES-like, where the uncertainty of $\sigma(s^\mathrm{G}_5)$ actually increases when combining with SKAO compared to MeerKLASS. 
			This counter-intuitive result stems from the \rep{acute sensitivity of the multi-tracer method to tomographic bin overlap. Specifically, the DES-like bin centred at $z_c^{DES} = \{0.90\}$ aligns more closely with the MeerKLASS bin, $z_c^{MK} = \{0.90\}$, than with SKAO bins, $z_c^{SKAO} = \{ 0.80, 1.00\}$. This discrepancy is sufficient for MeerKLASS to yield a better performance than SKAO. To understand this sensitivity, we tested an alternative SKAO binning, $z_c^{SKAO} = \{ 0.45, 0.65, 0.85, 1.05, 1.25\}$, which successfully resolved the seeming inconsistency.}
			This highlights the importance of considering \rep{redshift} overlap when designing multi-tracer strategies.

		\subsection{\rep{Impact of cross-tracer information}}\label{sec:Cross-Tracer_info}
	
\begin{figure}
				\centering
				\includegraphics[width=1\linewidth]{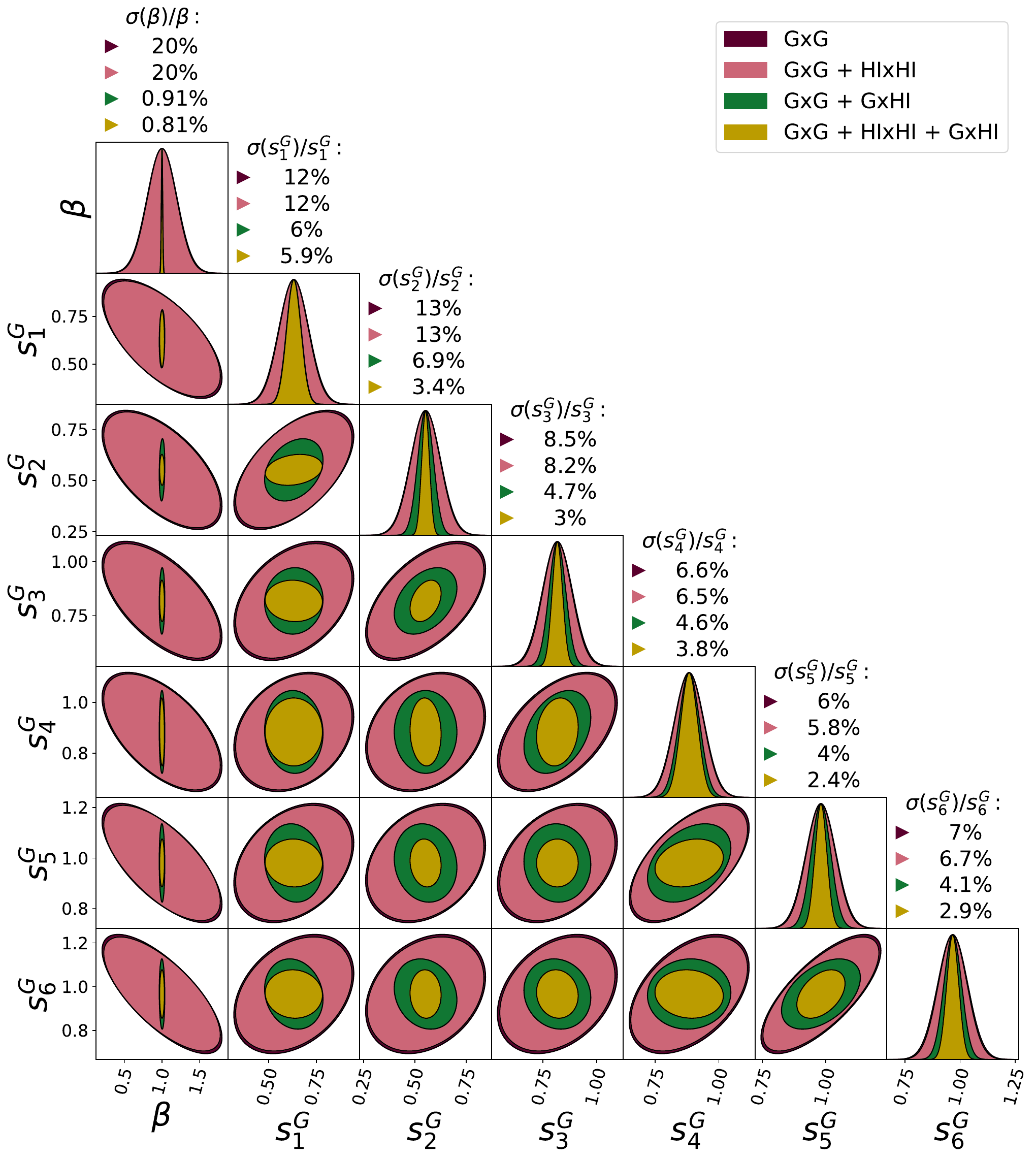}
				\caption{\rep{Corner plot showing the constraints on $\beta$ and $s^\mathrm{G}(z_i)$ for the DES-like and SKAO data combinations, comparing the four correlation scenarios outlined in \autoref{sec:Cross-Tracer_info}.}}
				\label{fig:CornerP_DESxSKA_GHI_cross}
			\end{figure}

			\rep{In \autoref{fig:CornerP_DESxHI} we observe a substantial improvement in the constraining power when doing a multi-tracer analysis, especially on $\beta$. 
			This enhancement can arise either from the H\textsc{i} IM autocorrelations, which help constrain the matter density and RSD components of the transfer function, or from the cross-correlations, which break degeneracies between $\beta$ and the magnification biases. 
			To isolate the dominant contribution, we compute one Fisher matrix for each of the following sets of $C_\ell$s:
			\begin{enumerate} 
				\item Galaxy only ($\mathrm{GxG}$);
				\item Galaxy correlations and H\textsc{i} correlations ($\mathrm{GxG + H\textsc{i}xH\textsc{i}}$);
				\item Galaxy correlations and cross-tracer correlations with H\textsc{i} (GxG + GxH\textsc{i});
				\item All correlations (GxG + H\textsc{i}xH\textsc{i} + GxH\textsc{i}).
			\end{enumerate}	
			The results for the DES-like and SKAO configurations are shown in \autoref{fig:CornerP_DESxSKA_GHI_cross}. 
			Comparing these scenarios reveals that the cross-tracer correlations between the two surveys provide the largest improvement in parameter constraints. 
			This indicates that the $\mathrm{GxH\textsc{i}}$ cross-information is the primary driver behind the enhancement of constraints on $\beta$ and $s^\mathrm{G}(z)$, by breaking degeneracies between them.}
		
	\section{Conclusions} \label{sec:Conclusions}
	
		In this work, we investigated synergies between photometric galaxy surveys and upcoming radio surveys, with the goal of constraining the weak lensing convergence and, ultimately, the gravitational potentials that encode the theory of gravity on cosmological scales.
		Specifically, we presented Fisher matrix forecasts for the magnification bias $s^\mathrm{G}(z)$ and the general relativistic probe parameter $\beta$.
		We demonstrated that combining the auto and cross-correlations of a galaxy survey with an H\textsc{i} intensity mapping survey via the multi-tracer method yields dramatically improved constraints on both parameters.

		First, we established the theoretical framework connecting weak lensing convergence to the metric potentials $\Phi$ and $\Psi$, introducing the parameter $\beta$ as a probe of modified gravity. Using this framework, we implemented Fisher matrix forecasts for three photometric galaxy surveys (DES-like, LSST-like, Euclid-like) and two H\textsc{i} intensity mapping surveys (MeerKLASS, SKAO).
		
		Then, we focused on forecasting constraints on parameters $\beta$ and $s^\mathrm{G}(z)$.
		Of the three galaxy surveys we tested, the one which gave better results with the multi-tracer method was the Euclid-like survey.
		This is due to multiple factors: being the only survey with tabulated values for $n_\mathrm{gal}$, $b^\mathrm{G}$ and $s^\mathrm{G}$ (\autoref{fig:Galaxy_Surveys-bz_ngal_sz}), having access to more redshift bins, and spanning a higher redshift range (\autoref{tab:z_c_values}).
		In \citet{Montanari_2015}, it was shown that including the cross-correlations of multiple redshift bins in a galaxy survey can constrain $\beta$ to a relative uncertainty of $1\%$. This value matches the result we obtained for a Euclid-like experiment. 
		However, by applying the multi-tracer method, we improved on this result, achieving a factor-of-eighteen reduction in uncertainty, down to \rep{$\sim0.01\%$}, highlighting the relevance of this method.
		Moreover, a comparison of our results from the DES-like survey with \citet{Legnani_2026_DES_Y6} indicates consistency with our findings: their constraints on $C_\mathrm{sample}$ (which corresponds to our $s^\mathrm{G}(z)$) have uncertainties ranging from $\sim2.5\%$ up to $\sim7.5\%$.
		Their range of uncertainties matches three of the six results we obtained for $s^\mathrm{G}(z)$; see \autoref{fig:CornerP_DESxHI}.
		
		Additionally, we verified that constraining $s^\mathrm{G}(z)$ and $\beta$ simultaneously does not pose a problem. 
		Although these parameters are locally degenerate (degenerate in autocorrelation measurements), the inclusion of tomographic data and its cross-correlations successfully starts to break down the degeneracy (see \autoref{fig:Histogram_Compare_results_sz_beta_DES_MK}).
		
		\rep{Extending this analysis, we further investigated which source of information yields the greatest improvement in the final constraints. Our tests indicate that the cross-tracer correlations are the primary driver of this enhancement, providing the largest gain in overall performance (see \autoref{fig:CornerP_DESxSKA_GHI_cross}).}
		
		The physical significance of these constraints lies in their connection to fundamental gravitational physics. In the standard $\Lambda$CDM model, $\beta = 1$; any deviation would signal a modification to general relativity, making $\beta$ a powerful consistency test for the cosmological standard model. Furthermore, precise measurements of $s^\mathrm{G}(z)$ allow us to isolate the lensing contribution to galaxy number counts $\Delta^L(z) = -(2-5s^\mathrm{G}(z))\kappa$, thereby constraining the convergence $\kappa$. In this way, we gain direct access to the sum of gravitational potentials $\Phi+\Psi$ through \autoref{eq:Mag_Conv_final}, a key observable for testing gravity on cosmological scales. The main result of this paper is that adding the radio information to photometric surveys opens a new window to constrain the amplitude of the Weyl potential. As seen in Figures \ref{fig:CornerP_DESxHI}, \ref{fig:CornerP_LSSTxHI}, and \ref{fig:CornerP_EuclidxHI} the forecast constraints on $\beta$ improve by an order of magnitude. 
		
		\rep{We validated our analysis by comparing results from the angular power spectrum, $C_\ell$, with those from a parallel pipeline based on the angular correlation function, $w(\theta)$, as detailed in \autoref{ap:ACF}. This comparison establishes $C_\ell$ as the preferred estimator for our forecasts. Even though $C_\ell$ only yields marginally tighter constraints than $w(\theta)$, its primary advantage lies in computational efficiency. The block-diagonal structure of its covariance matrix avoids the large, dense matrix inversions, required for $w(\theta)$, which become computationally prohibitive when multiple redshift bins are considered.}
		
		Several caveats and directions for future work merit discussion. First, our analysis assumes Gaussian covariance and does not account for non-linear structure formation at small scales, which may affect the highest multipoles considered. Second, we neglected systematic effects such as photometric redshift uncertainties beyond the simple scatter model, and foreground contamination in H\textsc{i} intensity mapping. A more complete treatment would incorporate these effects, though they are unlikely to qualitatively change our main conclusions. Third, the effectiveness of the multi-tracer method depends critically on overlapping \rep{redshift binning}; as we noted in the case of DES-like with SKAO, \rep{redshift bins with small overlap} can partially offset the gains from superior instrument specifications, highlighting the importance of survey coordination.
		
		A natural extension of this work would be to apply MCMC methods to actual or mock data, directly measuring $s^\mathrm{G}(z)$ and $\beta$. 
		Such an analysis could also incorporate additional probes, such as galaxy shear or CMB lensing, to further tighten the constraints on $\Phi+\Psi$. Looking further ahead, the combination of photometric surveys (Euclid) with next-generation radio arrays (SKAO) offers an unprecedented opportunity to test general relativity on cosmological scales with percent-level precision.

		In conclusion, we demonstrated that the multi-tracer combination of photometric galaxy surveys and H\textsc{i} intensity mapping surveys enables high-precision measurements of both the magnification bias $s^\mathrm{G}(z)$ and the gravity probe $\beta$. By exploiting their complementary systematics and overlapping sky coverage, this approach opens a clear pathway to constraining the combination $\Phi+\Psi$ and testing the validity of general relativity on the largest scales. As the next generation of cosmological surveys comes online, such multi-tracer synergies will prove essential for extracting the full scientific potential of these ambitious observational campaigns.

	\begin{acknowledgements}
		TS and JF thank the support of FCT - Fundação para a Ciência e a Tecnologia through national funds by these grants: UID/04434/2025 (DOI 10.54499/UID/04434/2025) and 2023.15069.PEX. JF acknowledges the support from FCT in the form of work through the Scientific Employment Incentive program (reference 2020.02633.CEECIND/CP1631/CT0002).
		
	\end{acknowledgements}

	\bibliographystyle{aa}
	\bibliography{References_Paper}

	\begin{appendix}
		\nolinenumbers
		
		\section{\rep{Consistent forecasts with the angular correlation function, $w(\theta)$}}\label{ap:ACF}
		
			A complementary statistic to the angular power spectrum is the two-point angular correlation function, denoted by $w(\theta)$, which also quantifies the clustering of galaxies but in an angular configuration space. 
			While $C_\ell$ operates in harmonic space, $w(\theta)$ offers an intuitive interpretation in real space \citep{Astro_Stats}. 
			This function describes the probability of having an excess of galaxy-pairs, at a given angular distance $\theta$, relative to the expected value (the mean density of galaxies).
			
			For this analysis we exploit the fact that both $C_\ell$ and $w(\theta)$ provide equivalent descriptions of the distribution of galaxies and are statistically related.  
			These two quantities are connected by the well-known relation \citep{Peebles_1980_LSS}
			\begin{equation}
				w^{\mathrm{W_X^{\rep{i}}W_{X'}^{\rep{j}}}}(\theta) =  \sum_{\ell \ge 0}^{\ell_{max}} \left( \frac{2\ell+1}{4\pi} P_\ell(\cos\theta) C^{\mathrm{W_X^{\rep{i}}W_{X'}^{\rep{j}}}}_{\ell}\right) \,\, ,
				\label{eq:w_theo}
			\end{equation}
			where \(P_\ell\) are the Legendre polynomials.	
			Here we introduced \(C^{\mathrm{W_X^{\rep{i}}W_{X'}^{\rep{j}}}}_{\ell}\) from \autoref{eq:C_ell_theo}, which gives us the tomographic expression of \(w(\theta)\) for the correlation between galaxies in redshift bins \( z_i \) and \( z_j \).
			This allows us to compute the angular correlation function directly from the \( C_\ell \) spectra obtained from our previous analysis. For this we use the publicly available code \texttt{PyCCL} \citep{2019ApJS..242....2C}.

			The expression for the Fisher matrix of either $w(\theta)$ can also be approximated by \autoref{eq:FisherM_general}.
			In it, we define the data vector as $\mathbf{D'} \equiv \vec w^{\rep{ij}}(\theta) = \{ w^\mathrm{G}(\theta), w^\mathrm{H\textsc{i}}(\theta), w^\mathrm{GH\textsc{i}}(\theta) \}$, and its covariance matrix as
			\begin{align}
				\mathbf C' = \text{Cov}\Big( \vec w^{\rep{ij}}(\theta), \vec w^{\rep{pq}}(\theta')\Big) \equiv \rep{\Gamma_{\theta\theta'}^{ijpq}}\,,
			\end{align}
			for which their expression was deduced in \autoref{ap:Cov_M_w_theta} and is given by \autoref{eq:CovM_w_final}.
			For the derivatives $\bar{\mathbf{D}'}_{,\mu} =\partial w^{\rep{ij}}(\theta)/ \partial \lambda_\mu$ see \autoref{ap:deriv_estimators}.
			
			Substituting by more explicit notation, we can rewrite the Fisher matrix expression of $w(\theta)$ as
			\begin{align}
				F_{\mu\nu} '
				\simeq \left( \frac{\partial \vec w^{\rep{ij}}(\theta)}{\partial \lambda_\mu} \right)^T \left(\rep{\Gamma_{\theta\theta'} ^{ijpq}}\right)^{-1}  \frac{\partial \vec w^{\rep{pq}}(\theta')}{\partial \lambda_\nu} \,.
				\label{eq:FisherM_w_general}	
			\end{align}

			Following the procedure to calculate the Fisher matrix, mentioned in \autoref{sec:FIM}, we obtain the relative uncertainty for each parameter $\lambda_\mu$ when calculated through the Fisher matrix of the $w(\theta)$ estimator.
			
\begin{figure}
				\centering
				\includegraphics[width=1.0\linewidth]{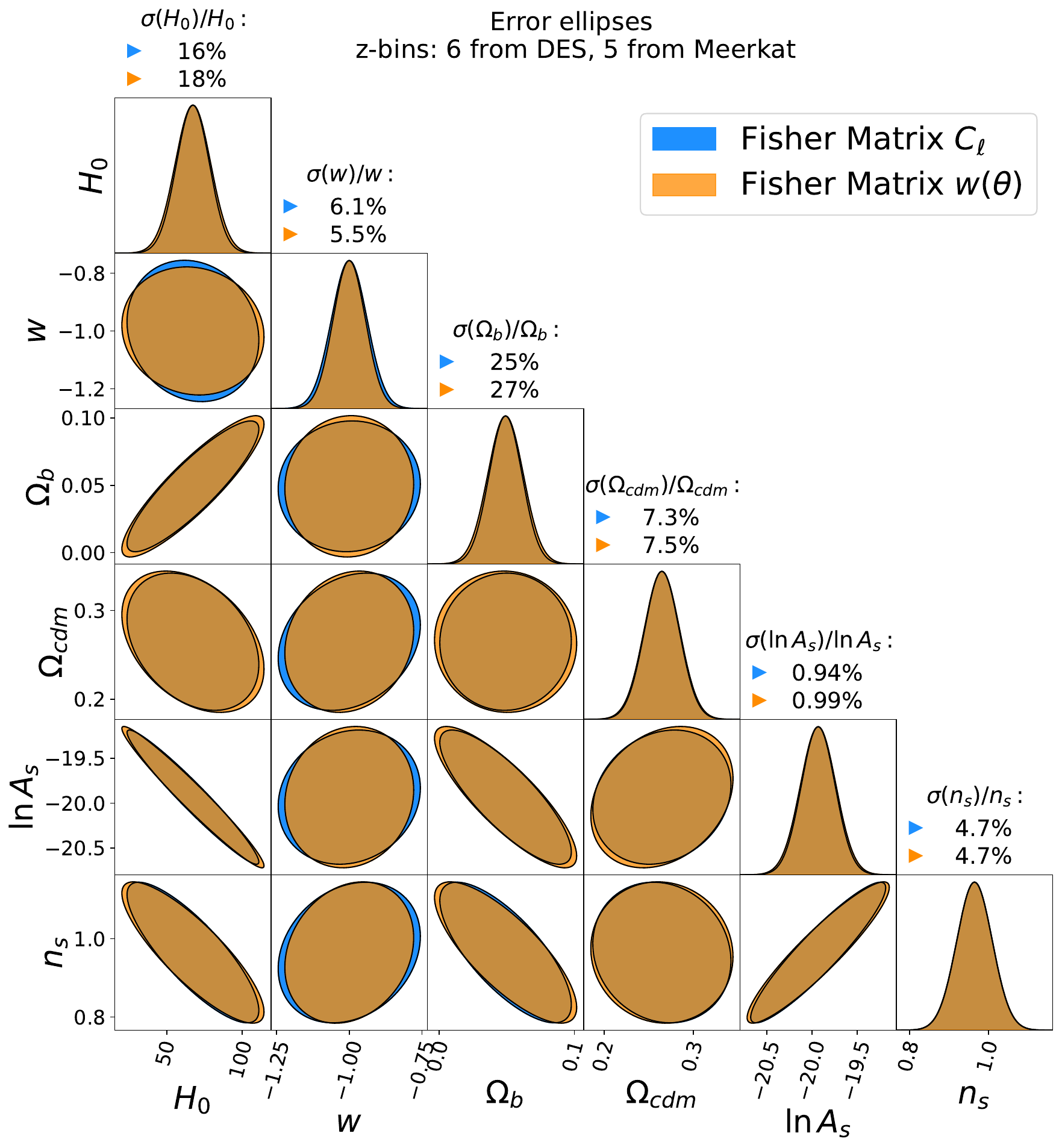}
				\caption{Corner plot showing the uncertainties of each cosmological parameter when constrained with data from the DES-like and MeerKLASS Surveys using $C_\ell$ and $w(\theta)$.}
				\label{fig:CornerP_DES_MK}
			\end{figure}

			A primary motivation for comparing $C_\ell$ and $w(\theta)$ stems from the fact that the Dark Energy Survey (DES) employs the angular correlation function \citep{Legnani_2026_DES_Y6}, while other surveys typically plan to use the angular power spectrum.
			In \autoref{fig:CornerP_DES_MK} we show this comparison through the error ellipses of the considered cosmological parameters, for the combination of a DES-like survey with MeerKLASS.
			The difference in performance between the two estimators is small, with $C_\ell$ yielding marginally tighter constraints across almost every parameter. 
			We verified also this conclusion for the parameters $s^\mathrm{G}(z)$ and $\beta$ across all survey combinations, finding consistent behaviour.

			\rep{However, it should be noted that since our analysis does not rely on real data, the values of the angular correlation function are derived directly from the angular power spectrum. Thus, the comparison between $w(\theta)$ and $C_\ell$ is limited and serves solely as a numerical validation tool. A proper comparison would take into account sky mask, weighting schemes, selection function and so on. Still,} practical considerations favour $C_\ell$. Unlike the block-diagonal covariance matrix of $C_\ell$, the covariance matrix of $w(\theta)$ requires computation for every entry. This down-weighs the potential advantage of $w(\theta)$ in dealing with the mask. When multiple redshift bin combinations are considered, we work with a large and dense covariance matrix that must be inverted to obtain the Fisher matrix. Such inversions of large matrices are computationally demanding and prone to numerical instability. Given these practical advantages of $C_\ell$, we decided to adopt it as the estimator of our analysis.

		\section{Expression for the covariance matrix of $C_\ell$} \label{ap:Cov_M_C_ell}
		
			The expression for the Gaussian covariance matrix of $\vec C_{\ell}^{\rep{ij}}$ is deduced in the appendix of \citet{Fonseca_2021} and reads
			\begin{align}
				\rep{\Gamma_{\ell\ell'} ^{ijpq}}
				&\equiv \text{Cov}\left(C_{\ell}^{\rep{ij}},C_{\ell'}^{\rep{pq}}\right) \nonumber\\
				&= \left\langle (C_{\ell}^{\rep{ij}} - \bar C_{\ell}^{\rep{ij}}) (C_{\ell'}^{\rep{pq}} - \bar C_{\ell'}^{\rep{pq}}) \right\rangle \nonumber\\
				&= \left\langle C_{\ell}^{\rep{ij}} C_{\ell'}^{\rep{pq}}\right\rangle  - \left\langle C_{\ell}^{\rep{ij}}\right\rangle \left\langle C_{\ell'}^{\rep{pq}} \right\rangle \nonumber\\
				&=\frac{\delta_{\ell\ell'}}{2\ell+1} \left[ C_{\ell}^{\rep{iq}}C_{\ell}^{\rep{pj}} + C_{\ell}^{\rep{ip}}C_{\ell}^{\rep{qj}} \right] \,. 
				\label{eq:CovM_Cell_general}
			\end{align}
			This expression is only valid if one has no noise in full sky. Once we introduce the noise, $\mathcal{N}^{\rep{ij}}$, either from the shot-noise in \autoref{eq:Shot_Noise_Gal} or the instrumental noise in \autoref{eq:Shot_Noise_HI}, the fraction of sky that its being observed, $f_{sky}$ and an $\ell$ step, $\Delta \ell$, one has
			\begin{equation}
				\begin{split}
					\rep{\Gamma_{\ell\ell'} ^{ijpq}} 
					= \frac{\delta_{\ell\ell'}}{2\ell+1} \frac{1}{f_\text{sky}\Delta\ell}
					&\Bigl[ \left(C_{\ell}^{\rep{iq}} + \mathcal{N}^{\rep{iq}}\right) \left(C_{\ell}^{\rep{pj}} + \mathcal{N}^{\rep{pj}}\right) \\
					+&  \left(C_{\ell}^{\rep{ip}} + \mathcal{N}^{\rep{ip}}\right) \left(C_{\ell}^{\rep{qj}} + \mathcal{N}^{\rep{qj}}\right) \Bigr]
					\,.
				\end{split} 
				\label{eq:CovM_Cell_noisy_lstep_ap}
			\end{equation}

		\section{Expression for the covariance matrix of $w(\theta)$}\label{ap:Cov_M_w_theta}
		
			For $w(\theta)$ we will write the theoretical covariance as a function of $C_\ell$. Neglecting the intrinsic noise contributions one has
			\begin{align*}
				\rep{\Gamma_{\theta\theta'}^{ijpq}} \equiv& \, \text{Cov}( w^{\rep{ij}}(\theta), w^{\rep{pq}}(\theta')) \\
				=& \, \left\langle (w^{\rep{ij}}(\theta) - \bar w^{\rep{ij}}(\theta))(w^{\rep{pq}}(\theta') - \bar w^{\rep{pq}}(\theta')) \right\rangle \nonumber\\
				=& \Bigg\langle \left[ \sum_\ell \frac{2\ell+1}{4\pi} P_\ell(\cos\theta) \left(C_{\ell}^{\rep{ij}} - \bar C_{\ell}^{\rep{ij}}\right) \right] \nonumber\\
				&\,\,\, \left[ \sum_{\ell'} \frac{2\ell'+1}{4\pi} P_{\ell'}(\cos\theta') \left(C_{\ell'}^{\rep{pq}} - \bar C_{\ell'}^{\rep{pq}}\right) \right] \Bigg\rangle \\
				=& \sum_\ell\sum_{\ell'} \bigg[ \frac{(2\ell+1)(2\ell'+1)}{(4\pi)^2}P_{\ell}(\cos\theta)P_{\ell'}(\cos\theta') \nonumber\\
				& \qquad\quad \left\langle \left(C_{\ell}^{\rep{ij}} - \bar C_{\ell}^{\rep{ij}}\right) \left(C_{\ell'}^{\rep{pq}} - \bar C_{\ell'}^{\rep{pq}}\right) \right\rangle \bigg] \,. 
			\end{align*}
			By substituting the covariance of \( \vec C_{\ell,ij}\) with \autoref{eq:CovM_Cell_general} we obtain
			\begin{align*}
				\rep{\Gamma_{\theta\theta'} ^{ijpq}} 
				=& \sum_\ell\sum_{\ell'} \bigg\{
				\frac{(2\ell+1)(2\ell'+1)}{16\pi^2}P_{\ell}(\cos\theta)P_{\ell'}(\cos\theta') \nonumber \\ 
				& \qquad\qquad \frac{\delta_{\ell\ell'}}{2\ell+1} \left[ C_{\ell}^{\rep{iq}}C_{\ell}^{\rep{pj}} + C_{\ell}^{\rep{ip}}C_{\ell}^{\rep{qj}} \right] \bigg\} \,. 
			\end{align*}
			This simplifies to the general expression of the covariance matrix of $w(\theta)$:
			\begin{equation}
				\begin{split}
					\rep{\Gamma_{\theta\theta'} ^{ijpq}} =& \sum_\ell \bigg\{ \frac{(2\ell+1)}{16\pi^2}P_{\ell}(\cos\theta)P_{\ell}(\cos\theta') \\ 
					& \qquad \left[ C_{\ell}^{\rep{iq}}C_{\ell}^{\rep{pj}} + C_{\ell}^{\rep{ip}}C_{\ell}^{\rep{qj}} \right] \bigg\} \,\,.
				\end{split} 
				\label{eq:CovM_w_general}
			\end{equation}

			Note that, for more specific cases, this general expression simplifies to more commonly used expressions (see \citet{Crocce_2011}), such as
			\begin{equation}
				\begin{split}
					\rep{\Gamma_{\theta\theta'} ^{ijij}} &= \sum_\ell \biggl\{ \frac{(2\ell+1)}{16\pi^2}P_{\ell}(\cos\theta)P_{\ell}(\cos\theta') \\
					& \qquad \quad \left[ \left(C_{\ell}^{\rep{ij}}\right)^2 + C_{\ell}^{\rep{ii}}C_{\ell}^{\rep{jj}} \right] \biggr\} \,,
				\end{split}
			\end{equation}
			and
			\begin{equation}
				\rep{\Gamma_{\theta\theta'} ^{iiii}} = \sum_\ell \biggl\{ \frac{(2\ell+1)}{16\pi^2}P_{\ell}(\cos\theta)P_{\ell}(\cos\theta') \left[ 2C_{\ell}^2 \right] \biggr\}\,\,.
			\end{equation}

			\autoref{eq:CovM_w_general} is already the general expression but, just like with \( C_\ell\) there are still two parameters missing: the noise, $\mathcal{N}$, and the fraction of the observed sky, \( f_{sky}\). 
			So, by introducing them in \autoref{eq:CovM_w_general}, we obtain
			\begin{align}
				\rep{\Gamma_{\theta\theta'} ^{ijpq}} = \frac{1}{f_{sky}}\sum_\ell \Bigg\{&
				\frac{(2\ell+1)}{16\pi^2}P_{\ell}(\cos\theta)P_{\ell}(\cos\theta') \nonumber\\ 
				&\Bigl[ \left(C_{\ell}^{\rep{iq}} + \mathcal{N}^{\rep{iq}}\right) \left(C_{\ell}^{\rep{pj}} + \mathcal{N}^{\rep{pj}}\right) \label{eq:CovM_w_noisy}\\
				+&  \left(C_{\ell}^{\rep{ip}} + \mathcal{N}^{\rep{ip}}\right) \left(C_{\ell}^{\rep{qj}} + \mathcal{N}^{\rep{qj}}\right) \Bigr] \Bigg\} \,. \nonumber
			\end{align}

			Moreover, when we calculate the covariance matrix of $w(\theta)$ we encounter a discretisation problem that doesn't appear in the $C_\ell$ analysis.
			This is due to the fact that, in the covariance matrix of $C_\ell$ we have defined values for each multipole $\ell$, since $\ell \in \mathbb N$.
			However, the angles have a wider range of possibilities, since $\theta \in \mathbb{R^+}$. This is problematic since, for a fixed range of values $\ell$, there is an equivalent range of angles that contain infinite values in it.
			Because of this, we will add a correction term to $\Gamma_{\theta\theta'}^{\rep{ijpq}}$, so that it accounts for this discretisation problem.
			We will create this term based on the correction added in the calculation of the mean.
			In other words, the same way that the mean of the data \( x \) has a discretisation correction of \( N \):
			\begin{equation*}
				\bar x = \frac{1}{N} \sum_i^N x^{\rep{i}} \,. 
			\end{equation*}
			We can also add this term to the covariance of \( x\) with \( y\):
			\begin{equation*}
				\begin{split}
					\text{Cov}(x, y) &= \Big\langle \big(x^{\rep{i}}-\bar x\big) \big(y^{\rep{i}} - \bar y\big) \Big\rangle \\
					&= \frac{1}{N} \sum_i^N \big(x^{\rep{i}}-\bar x\big) \big(y^{\rep{i}} - \bar y\big) \,,
				\end{split}
			\end{equation*}
			which, for \( x=w(\theta)\) and \( y=w(\theta')\) gives us 
			\begin{align*}
				\text{Cov}\big(w(\theta), w(\theta')\big) &= \Big\langle \big[w^{\rep{i}}(\theta)-\bar w(\theta)\big] \big[w^{\rep{i}}(\theta') - \bar w(\theta')\big] \Big\rangle \nonumber\\
				&= \frac{1}{N_\theta} \sum_i^N \bigg\{\big[w^{\rep{i}}(\theta)-\bar w(\theta)\big]
				\big[w^{\rep{i}}(\theta') - \bar w(\theta')\big] \bigg\} \,. \nonumber
			\end{align*}
			Here, \( N_\theta\) represents the number of bins that contain the angular information.
			So, its value is the separation between the central value of each bin, \( \Delta\theta\), divided by a scaling factor, which in this case will be the minimum value of \(\theta \) considered: 
			\begin{equation*}
				N_\theta = \frac{\Delta\theta}{\theta_{min}} \,.
			\end{equation*}
			
			Therefore, the transformation that we will apply to \autoref{eq:CovM_w_noisy} is
			\begin{equation*}
				\rep{\Gamma_{\theta\theta'} ^{ijpq}} \longrightarrow \frac{1}{N_{\theta}} \rep{\Gamma_{\theta\theta'} ^{ijpq}} \,,
			\end{equation*}
			which finally gives us the expression of the Covariance Matrix that we will use in our analysis:
			\begin{align}
				\rep{\Gamma_{\theta\theta'} ^{ijpq}} = \frac{1}{N_\theta f_{sky}}\sum_\ell &\Bigg\{
				\frac{(2\ell+1)}{16\pi^2}P_{\ell}(\cos\theta)P_{\ell}(\cos\theta') \nonumber\\ 
				&\Bigl[ \left(C_{\ell}^{\rep{iq}} + \mathcal{N}^{\rep{iq}}\right) \left(C_{\ell}^{\rep{pj}} + \mathcal{N}^{\rep{pj}}\right) \label{eq:CovM_w_final} \\
				+&  \left(C_{\ell}^{\rep{ip}} + \mathcal{N}^{\rep{ip}}\right) \left(C_{\ell}^{\rep{qj}} + \mathcal{N}^{\rep{qj}}\right) \Bigr]\Bigg\} \,\,. \nonumber
			\end{align}

		\section{Derivatives of both LSS estimators} \label{ap:deriv_estimators}
		
			There are two ways to calculate the derivatives of $C_\ell$ and $w(\theta)$: analytic derivatives and numerical ones. 
			Analytic derivatives are faster computationally, but they require a useful analytic expression.
			On the other hand, numerical derivatives take longer to compute but we don't need to know their expression.
			So, the use of analytic or numerical derivatives will depend on the parameter, $\lambda_\mu$, that we are considering. 
			
			The derivatives of $C_\ell$ in order to parameters $\beta$ and $A_s$ are calculated through the analytic expressions deduced in \autoref{ap:Anal_deriv_beta} and \autoref{ap:Anal_deriv_As}, respectively.
			All the other parameters will have to be solved numerically.
			For that we will use the five-point stencil approximation, which applied to $C_{\ell}^{\rep{ij}}$ results in: 
			\begin{equation}
				\resizebox{0.48\textwidth}{!}{$
					\frac{\partial C_{\ell}^{\rep{ij}}}{\partial \lambda_\mu} 
					\approx \frac {- C_{\ell}^{\rep{ij}}(\bar \lambda_\mu+2h_\mu)+8 C_{\ell}^{\rep{ij}}(\bar \lambda_\mu+h_\mu)-8 C_{\ell}^{\rep{ij}}(\bar \lambda_\mu-h_\mu)+ C_{\ell}^{\rep{ij}}(\bar \lambda_\mu-2h_\mu)}
					{12h_\mu} \,.$} 
				\label{eq:C_ell_five_point_stencil}
				\nonumber
			\end{equation}
			Here, we denoted $\bar \lambda_\mu$ as the fiducial value of the cosmological parameter that we are considering, and $h_\mu$ as the space between points. As a rule of thumb, the value of $h_\mu$, for a specific cosmological parameter $\mu$, is given by $ h_\mu \simeq 0,01 \ \bar \lambda_\mu$, although we used values where the numerical derivative was stable.

			Moving on to the derivatives of the $w(\theta)$ estimator, we can take advantage of \autoref{eq:w_theo} to find that
			\begin{align}
				\frac{\partial w^{\rep{ij}}(\theta)}{\partial \lambda_\mu} 
				&= \frac{\partial}{\partial \lambda_\mu}  \sum_{\ell \ge 0}^{\ell_{max}} \left( \frac{2\ell+1}{4\pi} P_\ell(\cos\theta) C_{\ell}^{\rep{ij}}\right) \nonumber\\			
				&= \sum_\ell \left(\frac{2\ell+1}{4\pi} P_\ell(\cos\theta) \left.\frac{\partial C_{\ell}^{\rep{ij}}}{\partial \lambda_\mu} \right|_{\lambda_\mu=\bar \lambda_\mu}\right) \,\,.
			\end{align}
			
			Therefore, in order to obtain the values of $\partial w^{\rep{ij}}(\theta) / \partial\lambda_\mu$ we can just use the values of $\partial C_{\ell}^{\rep{ij}}/ \partial \lambda_\mu$, which we already explored.

			\subsection{Analytic derivative of $ \partial C_\ell / \partial \beta $ } \label{ap:Anal_deriv_beta}

				First, note that the H\textsc{i} surveys don't have information on the lensing signal, implying that their derivatives of $ \partial C_\ell / \partial \beta $ are null.
				Then, by recalling \autoref{eq:C_ell_theo} and \autoref{eq:General_Frac_Pert_X_ell}, we can write the derivative of $C_\ell$, in order to $\beta$, as
				\begin{align*}
					\frac{\partial C^\mathrm{W_G^{\rep{i}}W_{G}^{\rep{j}}}_{\ell}}{\partial \beta} 
					&= \frac{\partial}{\partial\beta} \int \mathcal P(k) \Delta^\mathrm{W_{G}^{\rep{i}}}_{\ell} \Delta^\mathrm{W_{G}^{\rep{j}}}_{\ell} \ \dd{\ln k} \nonumber \\
					&= \frac{\partial}{\partial\beta} \int \mathcal P(k) \Bigg[\left(\Delta^\mathrm{W_M^{\rep{i}}}_{\ell} + \Delta^\mathrm{W_{RSD}^{\rep{i}}}_{\ell} + \Delta^\mathrm{W_{L}^{\rep{i}}}_{\ell}\right) \nonumber\\
					&\qquad\qquad\qquad\, \left(\Delta^\mathrm{W_{M}^{\rep{j}}}_{\ell} + \Delta^\mathrm{W_{RSD}^{\rep{j}}}_{\ell} + \Delta^\mathrm{W_{L}^{\rep{j}}}_{\ell}\right)\Bigg] \ \dd{\ln k} \nonumber \\
					&= \int \mathcal P(k) \frac{\partial}{\partial\beta} \left[ (\Delta^\mathrm{W_{M}}_\ell)^2 + (\Delta^\mathrm{W_{RSD}}_\ell)^2 + \Delta^\mathrm{W_{M}^{\rep{i}}}_{\ell}\Delta^\mathrm{W_{RSD}^{\rep{j}}}_{\ell}   \right. \\
					&\qquad \qquad \qquad 
					+ \Delta^\mathrm{W_{RSD}^{\rep{i}}}_{\ell}\Delta^\mathrm{W_{M}^{\rep{j}}}_{\ell} + (\Delta^\mathrm{W_{M}^{\rep{i}}}_{\ell} + \Delta^\mathrm{W_{RSD}^{\rep{i}}}_{\ell})\Delta^\mathrm{W_{L}^{\rep{j}}}_{\ell} \nonumber\\
					&\qquad \qquad \qquad \left.+ \Delta^\mathrm{W_{L}^{\rep{i}}}_{\ell}(\Delta^\mathrm{W_{M}^{\rep{j}}}_{\ell} \Delta^\mathrm{W_{RSD}^{\rep{j}}}_{\ell}) + (\Delta^\mathrm{W_{L}}_\ell)^2 \right] \ \dd{\ln k} \,.
					\nonumber
				\end{align*}
				So, when we derivate them in order of $\beta$, the terms with no lensing dependency will be null.
				The expression is then reduced to
				\begin{align}
					\frac{\partial C^\mathrm{W_G^{\rep{i}}W_{G}^{\rep{j}}}_{\ell}}{\partial \beta} 
					=& \int  \mathcal P(k) \frac{\partial}{\partial\beta} \Big[ (\Delta^\mathrm{W_{M}^{\rep{i}}}_{\ell} + \Delta^\mathrm{W_{RSD}^{\rep{i}}}_{\ell})\Delta^\mathrm{W_{L}^{\rep{j}}}_{\ell}  + \Delta^\mathrm{W_{L}^{\rep{i}}}_{\ell}(\Delta^\mathrm{W_{M}^{\rep{j}}}_{\ell} + \Delta^\mathrm{W_{RSD}^{\rep{j}}}_{\ell}) \nonumber\\
					& \qquad\qquad\quad  + (\Delta^\mathrm{W_{L}}_\ell)^2 \Big] \ \dd{\ln k} \,.
					\label{eq:ap_deriv_C_ell_beta_simp}
				\end{align}
				This expression can be further simplified by remembering that,  of the three components of $\Delta^\mathrm{W_{G}}$, only the Lensing component has the term $\beta$, recall  \autoref{eq:Delta_X_Lensing},
				\begin{equation*}
					\Delta^\mathrm{W_{L}^{\rep{i}}}_{\ell} = \ell(\ell+1)\left(1-\frac{5}{2}s^{X \rep{i}}\right)
					\int_{0}^{r} \frac{r-\tilde r}{r\tilde r} \beta \left(\tilde\Phi_{k}+\tilde\Psi_{k}\right)\;j_{\ell}(k\tilde r) \  \dd\tilde r \,,
					\label{eq:Delta_L_beta}
				\end{equation*} 
				and, since $\beta$ is a constant we can rewrite this equation around it, by defining the rest of the expression as $K$:
				\begin{equation*}
					\Delta^\mathrm{W_{L}^{\rep{i}}}_{\ell} = \beta  K_{\ell}^{\rep{i}} \,.
				\end{equation*}
				
				If we now substitute this expression into \autoref{eq:ap_deriv_C_ell_beta_simp}, we obtain
				\begin{align*}
					\frac{\partial C^\mathrm{W_G^{\rep{i}}W_{G}^{\rep{j}}}_{\ell}}{\partial \beta} 		
					&= \int \mathcal P(k) \frac{\partial}{\partial\beta} 
					\Bigg[ \beta  K^{\rep{j}}_{\ell} \left( \Delta^\mathrm{W_M^{\rep{i}}}_{\ell} + \Delta^\mathrm{W_{RSD}^{\rep{i}}}_{\ell}\right)  + \beta  K^{\rep{i}}_{\ell} \left(\Delta^\mathrm{W_M^{\rep{j}}}_{\ell} + \Delta^\mathrm{W_{RSD}^{\rep{j}}}_{\ell} \right) \nonumber \\
					& \qquad \qquad \qquad  
					+\beta^2  K^{\rep{i}}_{\ell}K^{\rep{j}}_{\ell}
					\Bigg] \ \dd{\ln k} \nonumber \\
					&= \int \mathcal P(k)
					\Bigg[  K^{\rep{j}}_{\ell} \left( \Delta^\mathrm{W_M^{\rep{i}}}_{\ell} + \Delta^\mathrm{W_{RSD}^{\rep{i}}}_{\ell}\right)  +  K^{\rep{i}}_{\ell} \left(\Delta^\mathrm{W_M^{\rep{j}}}_{\ell} + \Delta^\mathrm{W_{RSD}^{\rep{j}}}_{\ell} \right) \nonumber \\
					& \qquad \qquad \quad  
					+2\beta  K^{\rep{i}}_{\ell}K^{\rep{j}}_{\ell}
					\Bigg] \ \dd{\ln k} \, .\nonumber 
				\end{align*}

				Considering that we are using the $\Lambda \text{CDM}$ model, the fiducial value of $\beta$ is 1, so we can simplify our expression to
				\begin{align*}
					\frac{\partial C^\mathrm{W_G^{\rep{i}}W_{G}^{\rep{j}}}_{\ell}}{\partial \beta} 
					&= \int \mathcal P(k)
					\Bigg[  K^{\rep{j}}_{\ell} \left( \Delta^\mathrm{W_M^{\rep{i}}}_{\ell} + \Delta^\mathrm{W_{RSD}^{\rep{i}}}_{\ell}\right)  +  K^{\rep{i}}_{\ell} \left(\Delta^\mathrm{W_M^{\rep{j}}}_{\ell} + \Delta^\mathrm{W_{RSD}^{\rep{j}}}_{\ell} \right) \nonumber \\
					& \qquad \qquad \quad  
					+2  K^{\rep{i}}_{\ell}K^{\rep{j}}_{\ell}
					\Bigg] \ \dd{\ln k} \nonumber \\
					&= \int \mathcal P(k) \Bigl[ (\Delta^\mathrm{W_M^{\rep{i}}}_{\ell} + \Delta^\mathrm{W_{RSD}^{\rep{i}}}_{\ell})\Delta^\mathrm{W_L^{\rep{j}}}_{\ell} + \Delta^\mathrm{W_L^{\rep{i}}}_{\ell} (\Delta^\mathrm{W_M^{\rep{j}}}_{\ell} + \Delta^\mathrm{W_{RSD}^{\rep{j}}}_{\ell}) \nonumber \\
					& \qquad \qquad \quad +2 \Delta^\mathrm{W_L^{\rep{i}}}_{\ell} \Delta^\mathrm{W_L^{\rep{j}}}_{\ell} \Bigr] \ \dd{\ln k}  \,.  
				\end{align*}
				
				It might seem we're going nowhere, but if we now use the simple trick of adding and subtracting the terms: $\Delta^\mathrm{W_M^{\rep{i}}}_{\ell} \Delta^\mathrm{W_M^{\rep{j}}}_{\ell}$, $\Delta^\mathrm{W_{RSD}^{\rep{i}}}_{\ell} \Delta^\mathrm{W_{RSD}^{\rep{j}}}_{\ell}$, $\Delta^\mathrm{W_M^{\rep{i}}}_{\ell}\Delta^\mathrm{W_{RSD}^{\rep{j}}}_{\ell}$ and $\Delta^\mathrm{W_{RSD}^{\rep{i}}}_{\ell}\Delta^\mathrm{W_M^{\rep{j}}}_{\ell}$,  then the expression becomes
				\begin{align}
					\frac{\partial C^\mathrm{W_G^{\rep{i}}W_{G}^{\rep{j}}}_\ell}{\partial \beta} 
					=& \int \mathcal P(k) \Big[ \Delta^\mathrm{W_G^{\rep{i}}}_{\ell}\Delta^\mathrm{W_G^{\rep{j}}}_{\ell} + \Delta^\mathrm{W_L^{\rep{i}}}_{\ell} \Delta^\mathrm{W_L^{\rep{j}}}_{\ell} - \Delta^\mathrm{W_M^{\rep{i}}}_{\ell} \Delta^\mathrm{W_M^{\rep{j}}}_{\ell} - \Delta^\mathrm{W_{RSD}^{\rep{i}}}_{\ell} \Delta^\mathrm{W_{RSD}^{\rep{j}}}_{\ell} \nonumber\\
					& \qquad\qquad - \Delta^\mathrm{W_M^{\rep{i}}}_{\ell}\Delta^\mathrm{W_{RSD}^{\rep{j}}}_{\ell} - \Delta^\mathrm{W_{RSD}^{\rep{i}}}_{\ell}\Delta^\mathrm{W_M^{\rep{j}}}_{\ell}  \Big]
					\ \dd{\ln k} \nonumber\\
					=& \int \mathcal P(k) \Bigl[ \Delta^\mathrm{W_G^{\rep{i}}}_{\ell}\Delta^\mathrm{W_G^{\rep{j}}}_{\ell} + \Delta^\mathrm{W_L^{\rep{i}}}_{\ell} \Delta^\mathrm{W_L^{\rep{j}}}_{\ell} \nonumber\\
					& \qquad\qquad - (\Delta^\mathrm{W_M^{\rep{i}}}_{\ell} + \Delta^\mathrm{W_{RSD}^{\rep{i}}}_{\ell})(\Delta^\mathrm{W_M^{\rep{j}}}_{\ell} + \Delta^\mathrm{W_{RSD}^{\rep{j}}}_{\ell})  \Bigr]
					\ \dd{\ln k} \nonumber\\
					=& \int \mathcal P(k)  \Delta^\mathrm{W_G^{\rep{i}}}_{\ell}\Delta^\mathrm{W_G^{\rep{j}}}_{\ell} \ \dd{\ln k} + \int \mathcal P(k) \Delta^\mathrm{W_L^{\rep{i}}}_{\ell} \Delta^\mathrm{W_L^{\rep{j}}}_{\ell} \ \dd{\ln k} \nonumber \\
					& - \int \mathcal P(k) \left(\Delta^\mathrm{W_M^{\rep{i}}}_{\ell} + \Delta^\mathrm{W_{RSD}^{\rep{i}}}_{\ell}\right)\left(\Delta^\mathrm{W_M^{\rep{j}}}_{\ell} + \Delta^\mathrm{W_{RSD}^{\rep{j}}}_{\ell}\right) \ \dd{\ln k} \nonumber
				\end{align}
				\begin{equation}
					\frac{\partial C^\mathrm{W_G^{\rep{i}}W_{G}^{\rep{j}}}_\ell}{\partial \beta} 
					= C_{\ell}^\mathrm{W_G^{\rep{i}}W_{G}^{\rep{j}}} 
					+ C_{\ell}^{\mathrm{W_L^{\rep{i}} W_L^{\rep{j}}}} 
					- C_{\ell}^\mathrm{W_{M+RSD}^{\rep{i}} W_{M+RSD}^{\rep{j}} }\,\,.
					\label{eq:Cl_anal_deriv_appendix}
				\end{equation}
				So, from \autoref{eq:Cl_anal_deriv_appendix} we can tell that, by calculating the values of: $C_{\ell}^\mathrm{W_G^{\rep{i}}W_{G}^{\rep{j}}}$ (which includes Matter density, RSD and Lensing); $C_{\ell}^{\mathrm{W_L^{\rep{i}} W_L^{\rep{j}}}}$ (which only has the Lensing term); and of $C_{\ell}^\mathrm{W_{M+RSD}^{\rep{i}} W_{M+RSD}^{\rep{j}} }$ (which includes both matter density and RSD terms), we have enough information to determine $\partial C_{\ell}^{\rep{ij}} / \partial \beta$ analytically.

			\subsection{Analytic derivative of $ \partial C_\ell / \partial A_s $ } \label{ap:Anal_deriv_As}
			
				From \autoref{eq:C_ell_theo} and \autoref{eq:Primordial_Pk}, one can write the angular power spectrum as
				\begin{align*}
					C^\mathrm{W_X^{\rep{i}}W_{X'}^{\rep{j}}}_{\ell} &= \int \mathcal P(k) \Delta^\mathrm{W_{X}^{\rep{i}}}_{\ell}(k) \Delta^\mathrm{W_{X'}^{\rep{j}}}_{\ell}(k) \ \dd{\ln k} \nonumber\\
					&= \int  A_s \left(\frac{k}{k_0}\right)^{n_s-1} \Delta^\mathrm{W_{X}^{\rep{i}}}_{\ell}(k) \Delta^\mathrm{W_{X'}^{\rep{j}}}_{\ell}(k) \ \dd{\ln k} \,.
				\end{align*}
				
				Therefore, the derivation of $ \partial C_\ell / \partial A_s $ can be simplified to
				\begin{align*}
					\frac{\partial C^\mathrm{W_X^{\rep{i}}W_{X'}^{\rep{j}}}_{\ell}}{\partial A_s}
					&= \frac{\partial}{\partial A_s} \int A_s \left(\frac{k}{k_0}\right)^{n_s-1} \Delta^\mathrm{W_{X}^{\rep{i}}}_{\ell}(k) \Delta^\mathrm{W_{X'}^{\rep{j}}}_{\ell}(k) \ \dd{\ln k} \\
					&= \int \left(\frac{k}{k_0}\right)^{n_s-1} \Delta^\mathrm{W_{X}^{\rep{i}}}_{\ell}(k) \Delta^\mathrm{W_{X'}^{\rep{j}}}_{\ell}(k)\, \frac{\partial A_s}{\partial A_s}  \ \dd{\ln k} \\
					&= \int \left(\frac{k}{k_0}\right)^{n_s-1} \Delta^\mathrm{W_{X}^{\rep{i}}}_{\ell}(k) \Delta^\mathrm{W_{X'}^{\rep{j}}}_{\ell}(k)  \ \dd{\ln k}
				\end{align*}
				\begin{equation}
					\frac{\partial C^\mathrm{W_X^{\rep{i}}W_{X'}^{\rep{j}}}_{\ell}}{\partial A_s} = \frac{C^\mathrm{W_X^{\rep{i}}W_{X'}^{\rep{j}}}_{\ell}}{ A_s}\,\,.
				\end{equation}
		
	\end{appendix}
	
\end{document}